\documentclass[journal,12pt,onecolumn,draftclsnofoot,]{IEEEtran}
\RequirePackage[T1]{fontenc}
\RequirePackage{multirow}
\RequirePackage{graphicx}
\RequirePackage{mathptmx}      % use Times fonts if available on your TeX system
\RequirePackage{flushend}
\RequirePackage[numbers,sort&compress]{natbib}
\RequirePackage[colorlinks,citecolor=blue,urlcolor=blue,linkcolor=blue]{hyperref}
\usepackage{lscape}
\usepackage{rotating, adjustbox}
\usepackage[cmintegrals]{newtxmath}
\usepackage[table]{xcolor}
\usepackage{tikz}
\usetikzlibrary{arrows,shapes,positioning,shadows,trees}
\usetikzlibrary{trees}

\begin{document}

\title{Authentication schemes for Smart Mobile Devices: Threat Models, Countermeasures, and Open Research Issues}

%\author{Mohamed Amine Ferrag\thanksref{e1,addr1}
 %       \and
  %      Leandros Maglaras\thanksref{e2,addr2,addr3} %etc.
   %             \and
%Abdelouahid Derhab \thanksref{e3,addr4}
%\and
 %       Helge~Janicke\thanksref{e4,addr2}
%}

%\thankstext{e1}{e-mail: ferrag.mohamedamine@univ-guelma.dz}
%\thankstext{e2}{e-mail: leandros.maglaras@dmu.ac.uk}
%\thankstext{e3}{e-mail: abderhab@ksu.edu.sa}
%\thankstext{e4}{e-mail: heljanic@dmu.ac.uk}

%\institute{Department of Computer Science, Guelma University, 24000 Guelma, Algeria\label{addr1}
 %         \and
 %         School of Computer Science and Informatics, De Montfort University, Leicester, UK\label{addr2}
 %         \and
 %         General Secretariat of Digital Policy, Athens, Greece\label{addr3}
  %        \and
  %        Center of Excellence in Information Assurance (CoEIA), King Saud University, Saudi Arabia\label{addr4}
%}
\author{Mohamed Amine Ferrag$^{1}$, Leandros Maglaras$^{2}$, Abdelouahid Derhab $^{3}$, Helge Janicke $^{2}$

\thanks{$^{1}$ Department of Computer Science, Guelma University, 24000, Algeria e-mail: mohamed.amine.ferrag@gmail.com, ferrag.mohamedamine@univ-guelma.dz}
\thanks{$^{2}$ School of Computer Science and Informatics, De Montfort University, Leicester, UK, , and also with General Secretariat of Digital Policy, Athens, Greece, e-mail:  leandros.maglaras@dmu.ac.uk}
\thanks{$^{3}$ Center of Excellence in Information Assurance (CoEIA), King Saud University, Saudi Arabia, e-mail: abderhab@ksu.edu.sa}
}

\maketitle

\begin{abstract}
This paper presents a comprehensive investigation of authentication schemes for smart mobile devices. We start by providing an overview of existing survey articles published in the recent years that deal with security for mobile devices. Then, we give a classification of threat models in smart mobile devices in five categories, including, identity-based attacks, eavesdropping-based attacks, combined eavesdropping and identity-based attacks, manipulation-based attacks, and service-based attacks. This is followed by a description of multiple existing threat models. We also provide a classification of countermeasures into four types of categories, including, cryptographic functions, personal identification, classification algorithms, and channel characteristics. Therefore, according to the countermeasure characteristic used and the authentication model, we categorize the authentication schemes for smart mobile devices in four categories, namely, 1) biometric-based authentication schemes, 2) channel-based authentication schemes, 3) factors-based authentication schemes, and 4) ID-based authentication schemes. In addition, we provide a taxonomy and comparison of authentication schemes for smart mobile devices in form of tables. Finally, we identify open challenges and future research directions.

\end{abstract}
\begin{IEEEkeywords}
Security, Authentication, Smart Mobile Devices, Biometrics, Cryptography
\end{IEEEkeywords}

\section{Introduction}
Mobile devices are going to take a central role in the Internet of Things era \cite{111}. Smart phones, assisted from the 5G technology that provides continuous and reliable connectivity \cite{100}, will soon be able to support applications across a wide variety of domains like homecare, healthcare, social networks, safety, environmental monitoring, e-commerce and transportation \cite{2}. Storage capabilities of mobile phones increase rapidly, and phones can today generate and store large amounts of different types of data. Modern capabilities of smart phones such as mobile payment \cite{108} and mobile digital signing \cite{109} of documents can help the digitalization of both the private and the public sector raising new security and privacy requirements \cite{110}.

\begin{figure}
\centering
\includegraphics[width=1\linewidth]{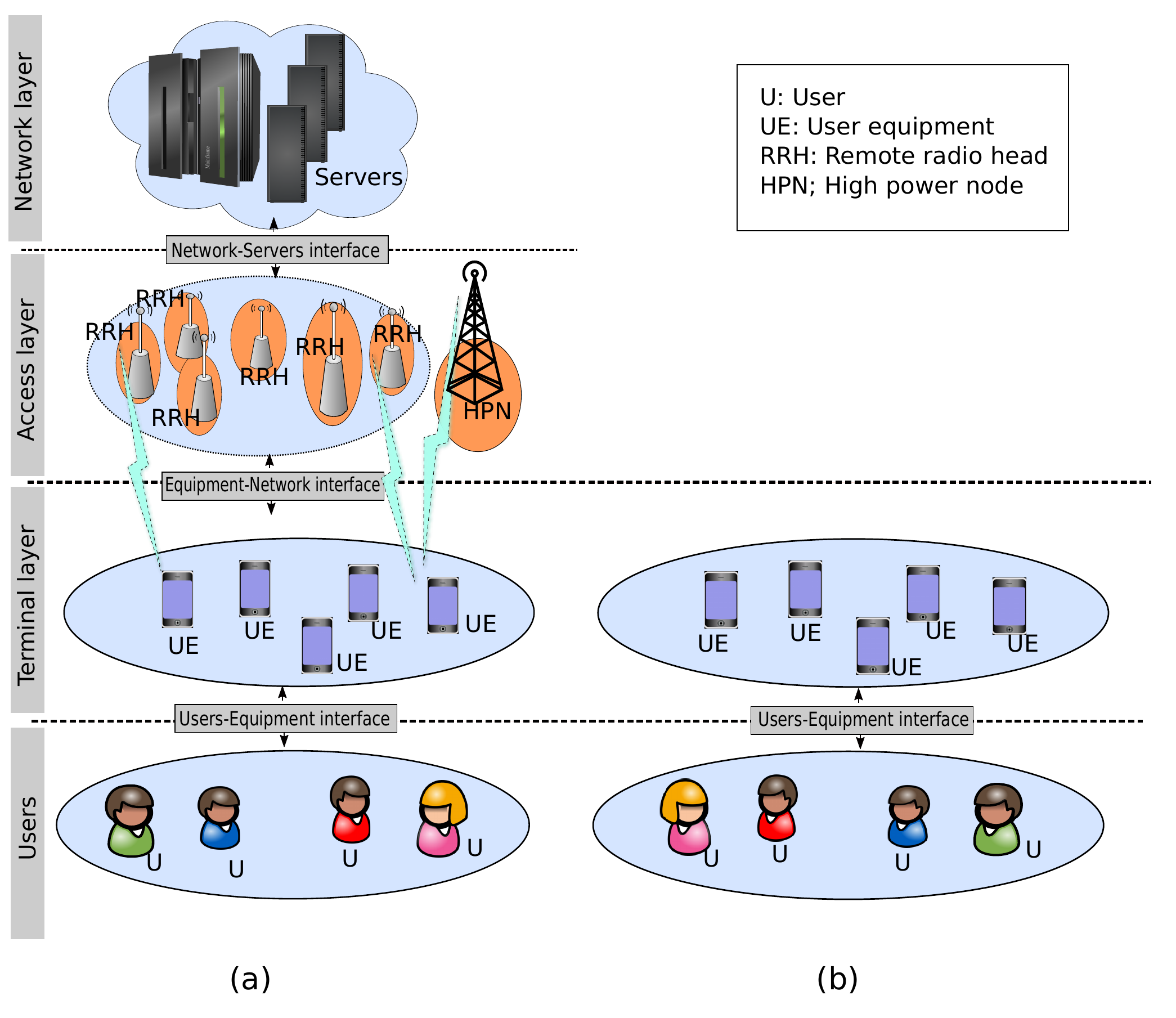}
\caption{Types of communication for the smart mobile devices during the authentication, (a) users accessing smart mobile devices, (b) users accessing remote servers via smart mobile devices}
\label{fig:Fig11}
\end{figure}

As shown in Figure \ref{fig:Fig11}, there are two types of access to smart mobile devices during the authentication phase, namely, 1) users accessing smart mobile devices, and 2) users accessing remote servers via smart mobile devices. Mobile devices are protected with the use of different methods ranging from single personal identification numbers PINs, passwords or patterns which have been proved to be vulnerable to different kinds of attacks \cite{3}. Moreover, it has been proven that the main breaches that systems face today, relate to attacks that can exploit human behavior, which cal for more sophisticated security and privacy measures \cite{112}. Even when strong authentication techniques are used during the initial access to the mobile device, there is a growing need for continuous authentication of legitimate users through users' physiological or behavioral characteristics \cite{10}. In this way, approaches, which exploit biometrics, like fingerprint recognition, face recognition, iris recognition, retina recognition, hand recognition or even dynamic behavior such as voice recognition, gait patterns or even keystroke dynamics, can help detect imposters in real time \cite{113}. Every new authentication method comes with a possible risk of low user acceptance due to latency and increasing complexity \cite{114}. 

In order to secure stored data from falling into wrong hands, cryptographic algorithms, which are conventional methods of authenticating users and protecting communication messages in insecure networks, can be used \cite{100}. Only the user who possesses the correct cryptographic key can access the encrypted content. Cryptographic algorithms can be categorized in two main groups \cite{18}, symmetric key cryptography and public key cryptography methods, where the latter although being more promising cannot be easily applied to short messages due to inducing big latency \cite{27}. In case an adversary obtains the secret key of a legitimate user, this kind of attack is very difficult to be detected in the server side.

To conduct the literature review, we followed the same process used in our previous work \cite{101}. Specifically, the identification of literature for analysis in this paper was based on a keyword search, namely, "authentication scheme", "authentication protocol", "authentication system", and "authentication framework". By searching these keywords in academic databases such as SCOPUS, Web of Science, IEEE Xplore Digital Library, and ACM Digital Library, an initial set of relevant sources were located. Firstly, only proposed authentication schemes for smart mobile devices were collected. Secondly, each collected source was evaluated against the following criteria: 1) reputation, 2) relevance, 3) originality, 4) date of publication (between 2007 and 2018), and 5) most influential papers in the field. The final pool of papers consists of the most important papers in the field of mobile devices that focus on the authentication as their objective. Our search started on 01/11/2017 and continued until the submission date of this paper. The main contributions of this paper are:

\begin{itemize}
\item We discuss the existing surveys on security for smart mobile devices.

\item We classify the threat models, which are considered by the authentication schemes in smart mobile devices, into five main categories, namely, identity-based attacks, eaves-dropping-based attacks, combined eavesdropping and identity-based attacks, manipulation-based attacks, and service-based attacks.

\item  We review existing research on countermeasures and security analysis techniques in smart mobile devices.

\item We provide a taxonomy and a side-by-side comparison, in a tabular form, of the state-of-the-art on the recent advancements towards secure and authentication schemes in smart mobile devices with respect to countermeasure model, specific mobile device, performance, limitations, computation complexity, and communication overhead.

\item We highlight the open research challenges and discuss the possible future research directions in the field of authentication in smart mobile devices.
\end{itemize}

The remainder of this paper is organized as follows.~Section \ref{sec:existing-surveys-on-security-for-smart-mobile-devices}~presents the existing surveys on security for mobile devices. In~Section \ref{sec:threat-models}, we provide a classification for the threat models for mobile devices. In Section \ref{sec:countermeasures-and-security-analysis-techniques}, we present countermeasures used by the authentication schemes for smart mobile devices. In~Section \ref{sec:authentication-schemes-for-smart-mobile-devices}, we present a side-by-side comparison in a tabular form for the current state-of-the-art of authentication schemes for mobile devices. Then, we discuss open issues and recommendations for further research in~Section \ref{sec:future-directions}. Finally, we draw our conclusions in~Section \ref{sec:conclusions}.

\begin{table*}[!t]
% * <d.ouahid@gmail.com> 2018-03-27T13:59:28.280Z:
% 
% > table
% In the table, some surveys are not listed:
% Aslam et al.
% Velasquez et al.
% Kilinc and Yanik
% 
% ^.
	\centering
	\caption{A summary of related survey papers}
	\label{Table:Tab1}
    {
\begin{tabular}{p{1.6in}|p{0.7in}|p{0.9in}|p{0.7in}|p{0.4in}|p{0.7in}|p{0.6in}} \hline \hline
\textbf{Ref.} & \textbf{Threat models} & \textbf{Countermeasures} & \textbf{Security analysis techniques} & \textbf{Security\newline Systems} & \textbf{Authentication schemes} & \textbf{Surveyed papers} \\ \hline \hline 
La Polla et al. (2013) \cite{1} & $\surd $ & $\surd $ & X & $\surd $ & 0 & 2004 - 2011 \\ \hline 
Khan et al. (2013) \cite{2} & X & X & X & $\surd $ & X & 2005 - 2008 \\ \hline 
Harris et al. (2014) \cite{8} & X & X & X & $\surd $ & X & 2005 - 2012 \\ \hline 
Meng et al. (2015) \cite{10} & 0 & $\surd $ & X & 0 & 0 & 2002 - 2014 \\ \hline 
Faruki et al. (2015) \cite{7} & $\surd $ & 0 & X & $\surd $ & X & 2010 - 2014 \\ \hline 
Teh et al. (2016) \cite{6} & X & 0 & X & 0 & 0 & 2012 - 2015 \\ \hline 
Alizadeh et al. (2016) \cite{5} & 0 & 0 & X & 0 & 0 & 2010 - 2014 \\ \hline 
Patel et al. (2016) \cite{3} & 0 & X & X & 0 & 0 & 2010 - 2015 \\ \hline 
Gandotra et al. (2017) \cite{4} & $\surd $ & 0 & X & $\surd $ & X & 2010 - 2015 \\ \hline 
Spreitzer et al. (2017) \cite{106} & $\surd $ & 0 & X & X & X & 2010 - 2016 \\ \hline 
Kunda and Chishimba (2018) \cite{200} & X & 0 & X & X & $\surd $ & 2010 - 2018 \\ \hline 
Our work\newline  & $\surd $ & $\surd $ & $\surd $ & $\surd $ & $\surd $ & 2007 - 2018 \\ \hline \hline
\end{tabular}
\\$\surd $ :indicates fully supported; X: indicates not supported; 0: indicates partially supported
}
\end{table*}

\section{Existing Surveys on Security for Smart Mobile Devices}\label{sec:existing-surveys-on-security-for-smart-mobile-devices}

There are around ten survey articles published in the recent years that deal with security for mobile devices. These survey articles are categorized as shown in Table \ref{Table:Tab1}. La Polla et al. in \cite{1} presented a survey on Security for Mobile Devices. They started by describing different types of mobile malware and tried to outline key differences between security solutions for smartphones and traditional PCs. They also presented the threats targeting smartphones by analyzing the different methodologies, which can be used to perform an attack in a mobile environment and explained how these methodologies can be exploited for different purposes. Based on their analysis, \textbf{ }which was conducted back in 2013, the authors present security solutions, focusing mostly on those that exploit intrusion detection systems and trusted platform technologies. In the same year, Khan et al.in \cite{2} performed a thorough survey on mobile devices, by considering them not as communication devices but as personal sensing platforms. Their research focused on two main categories, participatory and opportunistic mobile phone sensing systems. Having that in mind, they presented the existing work in the area of security of mobile phone sensing. They concluded that security and privacy issues need more attention while developing mobile phone sensing systems and applications, since as mobile phones are used for social interactions, users' private data are vulnerable. Harris et al. \cite{8} in their survey tried to identify all emerging security risks that mobile device imposes on SMEs and provided a set of minimum security recommendations that can be applied to mobile devices by the SMEs. Based on a fundamental dilemma, whether to move to the mobile era, which results in facing higher risks and investing on costly security technologies, or postpone the business mobility strategy in order to protect enterprise and customer data and information. 

Focusing on Android platforms, Faruki et al. in \cite{7} surveyed several security aspects, such as code transformation methods, strength, and limitations of notable malware analysis and detection methodologies. By analyzing several malware and different methods used to tackle the wide variety of new malware, they concluded that a comprehensive evaluation framework incorporating robust static and dynamic methods may be the solution for this emerging problem.  

Since password and PINs are authentication solutions with many drawbacks, Meng et al.  in \cite{10} conducted a thorough research on biometric-based methods for authentication on mobile phones. Authors included in their survey article both physiological and behavioral approaches, analyzed their feasibility of deployment on touch-enabled mobile phones, spotted attack points that exist and their corresponding countermeasures. Based on their analysis they concluded that a hybrid authentication mechanism that includes both multimodal biometric authentication along with traditional PINs or password can enhance both security and usability of the system.  In order to further enhance security and privacy of mobile devices, active authentication techniques, which constantly monitor the behavior of the user, are employed. These methods are surveyed in \cite{3}, where a thorough analysis of their advantages and limitations is presented along with open areas for further exploration. Using physiological and behavioral biometrics-based techniques similar to the ones surveyed in \cite{10} along with a continuous base and not only during initial access, but multimodal biometrics-based fusion methods have also been found to be the most efficient in terms of security and usability. One main issue that arises from the use of biometric characteristics is the possible theft of them, which can be prevented with the use of template protection schemes.  A similar survey \cite{6} that discusses touch dynamics authentication techniques for mobile devices was published in 2016. Touch dynamics is a behavioral biometrics, which captures the way a person interacts with a touch screen device both for static and dynamic authentication of users. Teh et al. in \cite{6} presented detailed implementations, experimental settings covering data acquisition, feature extraction, and decision-making techniques.

Alizadeh et al. in \cite{5} discussed authentication issues in mobile cloud computing (MCC) and compare it with that of cloud computing. They presented both Cloud-side and user authentication methods and spotted important that parameters that are important for designing modern authentication systems for MCC in terms of security, robustness, privacy, usability, efficiency, and adaptability. In another survey article that was published in 2017 \cite{106}, Spreitzer et al. focused on side-channel attacks against mobile devices and briefly discussed other attacks that have been applied in the smart card or desktop/cloud setting, since the interconnectivity of these systems makes smart phones vulnerable to them as well. Authors concluded that most of the attacks target Android devices, due to the big market share of Android platforms. They also recommended that future research should focus on wearables, e.g. smart watches, that may suffer from the same attacks in the near future, and pointed out that side-channel attacks can be combined with other attacks that exploit software vulnerabilities in order to be more efficient.

Aslam et al. in \cite{116} reviewed authentication protocols to access the Telecare Medical Information Systems and discussed their strengths and weaknesses in terms of ensured security and privacy properties, and computation cost. The schemes are divided into three broad categories of one-factor, two-factor, and three-factor authentication schemes. Velasquez et al. in \cite{118} presented existing authentication techniques and methods in order to discern the most effective ones for different contexts. In \cite{118}, Kilinc and Yanik reviewed and evaluated several SIP authentication and key agreement protocols according to their performance and security features. Finally in the last survey article, which was published in 2017 \cite{4}, Gandotra et al. surveyed device-to-device (D2D) communications along with security issues with the primary scope on jamming attacks.

From the above survey articles, only five of them deal with authentication schemes for mobile devices and none of them thoroughly covers the authentication aspects that are related to mobile devices. To the best of our knowledge, this work is the first one that thoroughly covers the aspects of: threat models, countermeasures, security analysis techniques, security systems, and authentication schemes that were recently proposed by the research community.

\begin{figure*}
\centering
	\begin{adjustbox}{width=0.9\textwidth}
		\begin{tikzpicture}[
		Level1/.style   = {rectangle, draw, rounded corners=1pt, thin, align=center, fill=blue!8},
		Level2/.style = {rectangle, draw, rounded corners=1pt, thin, align=center, fill=black!8, text width=8em},
		Level3/.style = {rectangle, draw, rounded corners=1pt, thin, align=left, fill=green!2, 	text width=7.8em},
		level 1/.style={sibling distance=35mm},
		edge from parent path={(\tikzparentnode.south)
			-- +(0,-8pt)
			-| (\tikzchildnode)}]
		
		\node[Level1] {Threat models}
		child {node[Level2] (c1) {Identity-based attacks}}
		child {node[Level2] (c2) {Eavesdropping-based attacks}}
		child {node[Level2] (c3) {Combined eavesdropping and identity-based attacks}}
		child {node[Level2] (c4) {Manipulation-based attacks}}
		child {node[Level2] (c5) {Service-based attacks}};

		\begin{scope}[every node/.style={Level3}, node distance=3mm]
		\node [below = of c1, xshift=0.5em] (c11) {\small{Deposit-key attack~\cite{30}}};
		\node [below = of c11] (c12) {\small{Impostor attack~\cite{12}}};
		\node [below = of c12] (c13) {\small{Impersonation attack~\cite{14,17,18,19,20,25,32,38}}};
		\node [below = of c13] (c14) {\small{Spoofing attack~\cite{14,24,38,44,68}}};
		\node [below = of c14] (c15) {\small{Masquerade attack~\cite{22,24}}};
		\node [below = of c15] (c16) {\small{Replay attack~\cite{17,18,19,25,26,27,30,32,38,21,69,22,44}}};
		\node [below = of c2, xshift=0.5em] (c21) {\small{Eavesdropping attack~\cite{69}}};
		\node [below = of c21] (c22) {\small{Adaptively chosen message attack~\cite{48}}};
		\node [below= of c22] (c23) {\small{Tracing attack~\cite{49}}};
		\node [below = of c23] (c24) {\small{Offline dictionary attack~\cite{38}}};
		\node [below = of c24] (c25) {\small{Outsider attack~\cite{18}}};
				\node [below = of c24] (c25) {\small{Outsider attack~\cite{18}}};
		\node [below = of c25] (c26) {\small{Brute force attack}};
		\node [below = of c26] (c27) {\small{Side-Channel attack~\cite{27}}};
		\node [below = of c27] (c28) {\small{Known-key attack~\cite{44}}};
		\node [below = of c28] (c29) {\small{Shoulder surfing and reflections~\cite{45}}};
		\node [below = of c29] (c210) {\small{Reflection attack~\cite{14,44}}};
		\node [below = of c210] (c211) {\small{Guessing attack~\cite{18,19,26,32}}};
		\node [below = of c211] (c212) {\small{ID attack~\cite{20, 23, 49}}};
		
		\node [below = of c3, xshift=0.5em] (c31) {\small{Malicious user attack}};
		\node [below = of c31] (c32) {\small{Parallel session attack}};
		\node [below= of c32] (c33) {\small{Stolen-verifier attack}};

		\node [below = of c4, xshift=0.5em] (c41) {\small{Man-in-the-middle (MITM) attack~\cite{44,21,12,27,41,57}}};
		\node [below = of c41] (c42) {\small{Forgery attack~\cite{14,21,69}}};
		\node [below = of c42] (c43) {\small{Trojan horse attack~\cite{27}}};
		\node [below = of c43] (c44) {\small{Biometric template attack~\cite{27}}};

		\node [below = of c5, xshift=0.5em] (c51) {\small{DDoS/DoS attack~\cite{14,30,32,44}}};
	
		\end{scope}
		
		\foreach \value in {1,2,3,4,5,6}
		\draw[-] (c1.195) |- (c1\value.west);
		
		\foreach \value in {1,...,10,11,12}
		\draw[-] (c2.195) |- (c2\value.west);

		\foreach \value in {1,2,3}
		\draw[-] (c3.195) |- (c3\value.west);
		
		\foreach \value in {1,2,3,4}
		\draw[-] (c4.195) |- (c4\value.west);
		
		\foreach \value in {1}
		\draw[-] (c5.195) |- (c5\value.west);
		\end{tikzpicture}
		
	\end{adjustbox}\caption{Classification of threat models in smart mobile devices}
\label{fig:Fig2a}
\end{figure*}
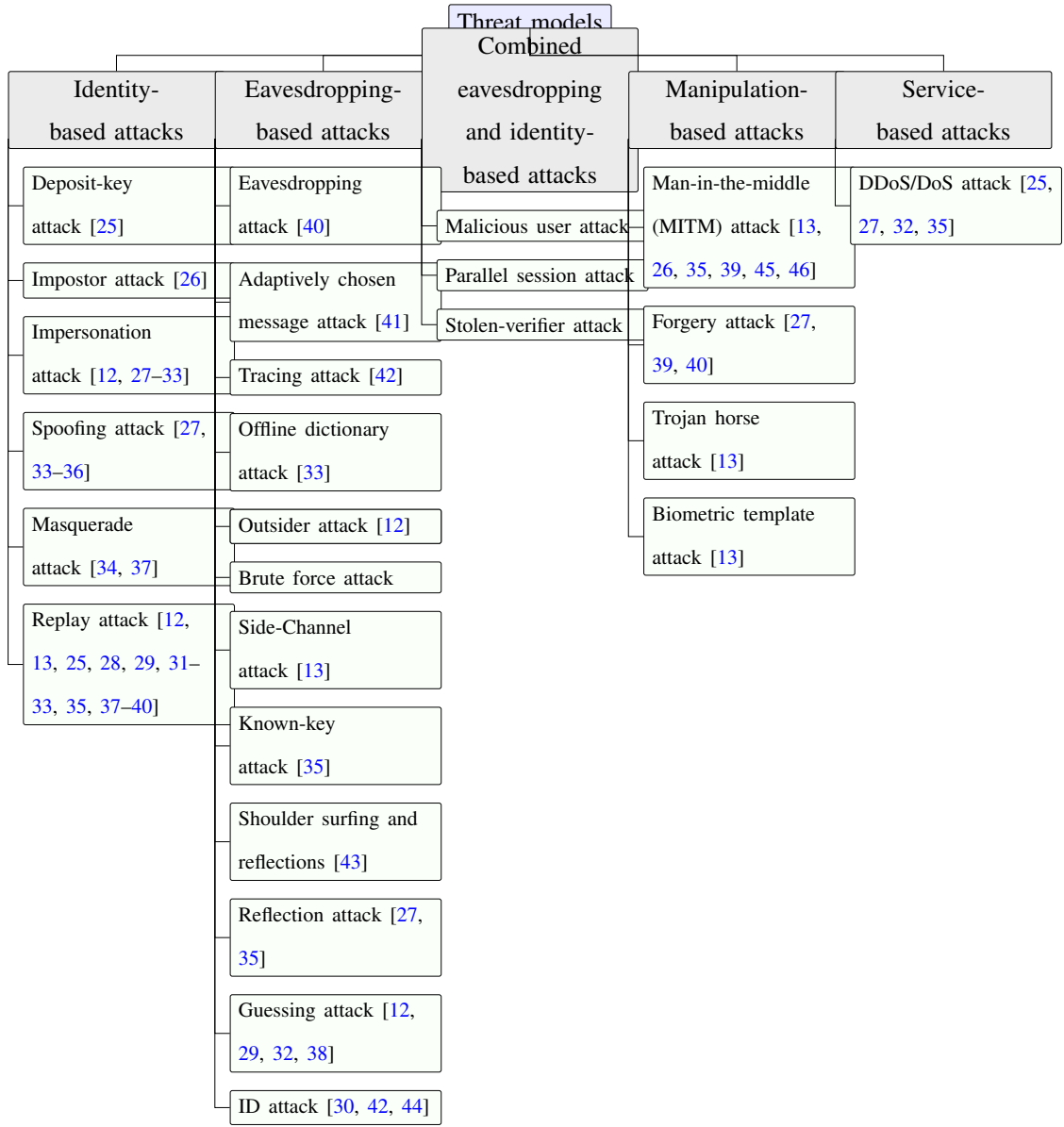

\section{Threat Models}\label{sec:threat-models}

In this section, we present and discuss the threat models that are considered by the authentication protocols in smart mobile devices. A summary of 26 attacks are classified into the following five main categories: identity-based attacks, eavesdropping-based attacks, Combined eavesdropping and identity-based attacks, manipulation-based attacks, and service-based attacks, as presented in Figure \ref{fig:Fig2a}.

\subsection{Identity-based attacks}
The attacks under this category forge identities to masquerade as authorized users, in order to get access to the system. We classify 6 attacks, namely: Deposit-key attack, Impostor attack, Impersonation attack,  Spoofing attack, Masquerade attack, and Replay attack.

\begin{itemize}

\item Deposit-key attack: It involves three parties: a roaming user, the user's home server, and the visiting foreign server of the roaming user. Under this attack, a malicious server makes the visiting foreign server believes that it is the user's home server. The roaming user deposits information at the visiting foreign server, which is also accessible by the user's fake home server (i.e., malicious server). In \cite{30}, this attack can be detected by verifying the key of foreign servers, the user can know that the foreign server does not think that its home server is the malicious server.

\item Impostor attack: An adversary disables one of the co-located devices and attempts to impersonate it. To thwart this attack, the Diffie-Hellman key exchange is extended with a co-location verification stage to ensure that the pairing takes place between two co-located devices \cite{12}.

\item Impersonation attack: An adversary tries to masquerade as a legitimate to log into the server. As presented in Figure \ref{fig:Fig2a}, there are eight authentication protocols \cite{14,17,18,19,20,25,32,38} that are resilient against this attack, and which use different ideas. The idea of chaotic hash-based fingerprint biometric is used in \cite{14}. The idea of asymmetric encryption function is used by \cite{17}. The idea of Elliptic curve cryptosystem is used in protocols \cite{18,19}. In addition, \cite{20} uses bilinear pairings, \cite{25} is based on an initial random seed number that is generated by the authorization authority. \cite{32} and \cite{38} adopt techniques based on Hashing functions and self-certified public keys respectively.

\item Spoofing attack: An adversary masquerades as a legal server to cheat a remote user. As shown in in Figure \ref{fig:Fig2a}, there are 5 authentication protocols \cite{14,24,38,44,68}, which are proposed to prevent and detect this attack. The idea of mutual authentication is used in protocols \cite{14,24,38}. \cite{44} is based on Key-hash based fingerprint remote authentication scheme. Besides, the pattern recognition approaches are adopted in \cite{68}.

\item Masquerade attack: An adversary may try to masquerade as a legitimate user to communicate with the valid system or a legitimate user. \cite{22} uses the idea of ransom values. An adversary cannot fabricate a fake request authentication message as it does not know the random value of a legitimate user and hence cannot masquerade as that user. On the other hand, the idea of mutual authentication is used in \cite{24}.

\item Replay attack: It consists of spoofing the identities of two parties, intercepting data packets, and relaying them to their destinations without modification.  As shown in Figure \ref{fig:Fig2a}, there are 13 authentication protocols \cite{17,18,19,25,26,27,30,32,38,21,69,22,44} to deal with this attack. The idea of using signatures during the authentication phase is proposed in \cite{17}. The idea of using different nonce variables in each login is adopted by protocols in \cite{32,38}. Protocols in \cite{18,19,25,26,27,30} use the idea of timestamps, which is combined with a randomly chosen secret key in protocols \cite{21,69}. On the other hand, \cite{22} proposes a one-way hashing function and random values, and \cite{44} proposes random nonce and three-way challenge-response handshake technique.

\end{itemize}

\subsection{Eavesdropping-based attacks}

This category of attacks is based on eavesdropping the communication channel between the user and the server in order to get some secret information and compromise the confidentiality of the system. We can list the following attacks under this category:

\begin{itemize}

\item Eavesdropping attack: An attacker secretly overhears information that is transmitted over the communication channel, and which might not be authorized to know. The protocol in \cite{17} deals with this attack by using the one-way hash function. On the other hand, the protocol in \cite{69} uses encryption with the pairwise master key.

\item Adaptively chosen message attack: Under this attack, an adversary attempts to forge a valid signature with the help of the private key generator (PKG). The objective of this attack is to gradually reveal information about an encrypted message or about the decryption key. To do so, ciphertexts are modified in specific ways to predict the decryption of that message. The protocol in \cite{48} can resist against this attack as it uses a certificateless signature.

\item Tracing attack: An adversary aims to collect enough privacy information to link data to a particular real identity. To resistant against this attack, \cite{49} uses random numbers in commitments and proofs.

\item Offline dictionary attack: An attacker collect useful information from the insecure channel or from the lost smart card. After that, he stores them locally and then uses them to generate a guessed password to verify the correctness of his guess. To thwart this attack, \cite{38} employs the password salting mechanism.

\item Outsider attack: An adversary uses the overhead messages that are exchanged between user and server, in order to compute the secret key of the server. This attack is prevented in \cite{18} by using the elliptic curve cryptosystem. 

\item Brute force attack: It consists of generating a large number of consecutive guessed passwords, with the hope of eventually guessing correctly. The resiliency against this attack is strengthened by employing the cryptographic hash function SHA-224. 

\item Side-Channel attack: It is based on information gained from the physical implementation of the cryptosystem. The physical electronic systems produce emissions about their internal process, which means that attackers can gather and extract cryptographic information. To resist against this attack,  \cite{27} proposes deploying elliptic curve cryptosystem as well as a Public Key Infrastructure (PKI).

\item Known-key attack: It consists of compromising past session keys in order to derive any further session keys. In \cite{44}, the values that are used to compute the session keys are not available in plaintext. In addition, random nonce imparts dynamic nature to the session key, and hence the attacker cannot predict the value of the random nonce of the future session key.

\item Shoulder surfing and reflections: It is a social engineering technique used to obtain information such passwords and other confidential data by looking over the victim's shoulder. To prevent this attack, \cite{45} uses the idea of sightless two-factor authentication.

\item Reflection attack: It is applicable to authentication schemes that adopt challenge-response technique for mutual authentication. Under this attack, a victim is tricked to provide the response to its own challenge. To deal with this attack, \cite{14} proposes the chaotic hash-based fingerprint biometric remote user authentication scheme, and \cite{44} proposes the key-hash based fingerprint remote authentication scheme.

\item Guessing attack: This attack is possible when an adversary gets a copy of the encrypted password from the communication channel or from the smart card. Then, the adversary guesses thousands of passwords per second and matches them with the captured one until the guessing operation succeeds. To deal with this attack, protocols \cite{18,19,26,32} use the elliptic curve cryptosystem.

\item ID attack: An adversary sends some identities to obtain the private key of the corresponding identity. The security against this attack is ensured in \cite{20, 23, 49} by using the idea of bilinear pairings.

\end{itemize}

\subsection{Combined Eavesdropping and identity-based attacks}

This category of attacks combines the eavesdropping and identity-based techniques to compromise the system. Under this category, we can find the following attacks: 

\begin{itemize}

\item Malicious user attack: An attacker by extracting the credentials stored in the smart card, he can easily derive the secret information of the system. After that, he masquerades as a legitimate user and accesses the system.

\item Parallel session attack: This attack takes place under the assumption that multiple concurrent sessions are allowed between two communicating parties. An attacker that eavesdrops over an insecure channel and captures login authentication message from the user and the responding authentication message from the server, can create and send a new login message to the server, and masquerading as the user.

\item Stolen-verifier attack: The attacker steals the verification data from the server of a current or past successful authentication session. Then, it uses the stolen data to generate authentication messages and send them to the server. If the server accepts the authentication messages, the adversary masquerades as a legitimate user.

\end{itemize}
\subsection{Manipulation-based attacks}

A data manipulation attack typically involves an unauthorized party accessing and changing your sensitive data, rather than simply stealing it or encrypting your data and holding it for ransom.

\begin{itemize}

\item Man-in-the-middle (MITM) attack: An attacker by spoofing the identities of two parties can secretly relay and even modify the communication between these parties, which believe they are communicating directly, but in fact, the whole conversation is under the control of the attacker. \cite{44} proposes the key-hash based fingerprint remote authentication scheme to secure the system against this attack. In \cite{12}, Diffie-Hellman key exchange with a co-location verification stage is proposed. \cite{21} combines bilinear pairing and elliptic curve cryptography. On the other hand, \cite{27} uses the idea of combining biometric fingerprint and the ECC public key cryptography, whereas symmetric encryption and message authentication code are used in \cite{41}. The Multi factors-based authentication scheme is adopted in \cite{57}.

\item Forgery attack: It forges valid authentication messages to satisfy the requirement of the authentication scheme. To resist against this attack, \cite{14} proposes the chaotic hash-based fingerprint biometric remote user authentication scheme. On the other hand, \cite{21,69} uses the idea of pairing and elliptic curve cryptography. 

\item Trojan horse attack: It uses a Trojan horse program to compromise the authentication system. In order to prevent that the Trojan horse program tampers with the biometric authentication module, \cite{27} seamlessly integrates biometric and cryptography.

\item Biometric template attack: An adversary attacks the biometric template in the database to add, modify, and delete templates in order to gain illegitimate access to the system. To increase the security strength of the biometric template, \cite{27} maximizes its randomness.

\end{itemize}

\subsection{Service-based attacks}

The goal of service-based, or Denial of Service (DoS) attacks, is to make the authentication service unavailable either by (1) flooding the authentication server with huge amount of data to make it busy and unable of providing service to the legitimate users, or (2) updating the verification information of a legitimate user with false data. Afterward, a legitimate legal user is unable to login to the server. As depicted in Figure \ref{fig:Fig2a}, there are four authentication protocols \cite{14,30,32,44} to prevent or detect DoS attacks. In \cite{14}, the user has to perform authentication by using a biometric fingerprint. If the mobile device is stolen or lost, illegitimate users cannot make a new password, and hence \cite{14} is resistant against the denial-of-service attack. As for protocol in \cite{30}, it is only required that the user and the foreign server to be involved in each run of the protocol, and the home server can be off-line. Consequently, DoS attack on home servers is not possible. On the other hand, \cite{32} uses the idea of one-way hash function, and \cite{44} proposes a key-hash based fingerprint remote authentication scheme.

\begin{table*}[!t]
	\centering
	\caption{Countermeasures used by the authentication schemes for smart mobile devices}
	\label{Table:Tab3}
{\scriptsize
\begin{tabular}{p{2.5in}|p{1.5in}} \hline \hline
\textbf{Countermeasure} & \textbf{Scheme} \\ \hline \hline 
Personal Identification Number (PIN) & \cite{11} \cite{13} \cite{43} \cite{57} \\ \hline 
Ear Shape & \cite{79} \\ \hline 
Electrocardiogram & \cite{61} \cite{74} \\ \hline 
Capacitive touchscreen & \cite{52} \\ \hline 
Behaviour profiling & \cite{58} \\ \hline 
Linguistic profiling & \cite{58} \\ \hline 
Gait recognition & \cite{51} \\ \hline 
Rhythm & \cite{50} \cite{54} \\ \hline 
Touch dynamics & \cite{33} \cite{45} \\ \hline 
Multi-touch interfaces & \cite{35} \cite{36} \\ \hline 
Probabilistic polynomial time algorithms & \cite{30} \\ \hline 
Initial random seed number & \cite{25} \\ \hline 
A unique international mobile equipment identification number & \cite{25} \\ \hline 
Encryption with pairwise master key & \cite{69} \\ \hline 
Identity-based elliptic curve algorithm & \cite{69} \\ \hline 
Tag number & \cite{17} \\ \hline 
Keystroke analysis & \cite{11} \cite{13} \cite{31} \cite{34} \cite{43} \cite{47} \cite{58} \cite{64} \\ \hline 
Diffie-Hellman key exchange & \cite{12} \\ \hline 
Classification algorithms & \cite{13} \cite{15} \cite{24} \cite{39} \cite{70} \\ \hline 
Chaotic hash & \cite{14} \\ \hline 
Fingerprint & \cite{14} \cite{27} \cite{32} \cite{44} \cite{75} \\ \hline 
Teeth image & \cite{15} \\ \hline 
Voice recognition & \cite{15} \cite{43} \cite{57} \\ \hline 
HMM biosensor scheduling & \cite{16} \\ \hline 
Asymmetric encryption function & \cite{17} \\ \hline 
Symmetric encryption function & \cite{17} \cite{24} \cite{25} \cite{41} \\ \hline 
Hash function & \cite{17} \cite{14} \cite{21} \cite{22} \cite{23} \cite{24} \cite{25} \cite{26} \cite{27} \cite{37} \cite{38} \cite{44} \cite{67} \cite{104} \cite{83} \\ \hline 
Elliptic curve cryptosystem & \cite{18} \cite{19} \cite{21} \cite{26} \cite{27} \cite{104} \cite{83} \\ \hline 
Bilinear pairings & \cite{20} \cite{21} \cite{23} \cite{37} \cite{38} \cite{49} \cite{67} \\ \hline 
Password & \cite{22} \cite{24} \cite{34} \cite{41} \cite{57} \\ \hline 
Schnorr's signature scheme & \cite{37} \\ \hline 
Self-certified public keys & \cite{38} \\ \hline 
Graphical password & \cite{40} \\ \hline 
Message authentication code & \cite{41} \\ \hline 
Channel characteristics & \cite{42} \\ \hline 
Face recognition & \cite{46} \cite{57} \cite{70} \\ \hline 
Iris recognition & \cite{46} \cite{68} \\ \hline 
Certificateless signature & \cite{48} \\ \hline 
Homomorphic encryption & \cite{49} \cite{56} \\ \hline 
Order preserving encryption & \cite{56} \\ \hline 
Gaze gestures & \cite{62} \\ \hline 
Arm gesture & \cite{79} \\ \hline 
Signature recognition  & \cite{90} \\ \hline \hline
\end{tabular}}
\end{table*}

\begin{table*}[!t]
	\centering
	\caption{Security analysis techniques used by the authentication schemes for smart mobile devices}
	\label{Table:Tab4}
{
\begin{tabular}{p{0.2in}|p{0.2in}|p{1.6in}|p{1.5in}|p{2.5in}} \hline \hline
\textbf{Ref.} & \textbf{Time} & \textbf{Tool} & \textbf{Authentication model} & \textbf{Main results} \\ \hline \hline  
\cite{13} & 2007 & - Pattern recognition approaches & - User authentication & - Evaluating the feasibility of utilizing keystroke information in classifying users \\ \hline 
\cite{15} & 2008 & - Pattern recognition approaches & - User authentication & - Evaluating the feasibility of utilizing together teeth image and voice \\ \hline
\cite{20} & 2009 & - Random oracle model\newline - Computational assumptions & - Mutual authentication & - Show that the proposed protocol is secure against ID attack \\ \hline 
\cite{21} & 2009 & - Computational assumptions & - Hand-off authentication\newline - Anonymous authentication & - Show that the proposed scheme can protecting identity privacy \\ \hline 
\cite{23} & 2010 & - Random oracle model\newline - Computational assumptions & - Mutual authentication & - Show that an adversary should not know the previous session keys \\ \hline 
\cite{33} & 2012 & - Pattern recognition approaches & - User authentication & - Evaluating the feasibility of touch dynamics \\ \hline 
\cite{35} & 2012 & - Pattern recognition approaches & - User authentication & - Show that the multi-touch gestures great promise as an authentication mechanism \\ \hline 
\cite{36} & 2012 & - Pattern recognition approaches & - Continuous mobile authentication & - Evaluating the applicability of using multi-touch gesture inputs for implicit and continuous user identification \\ \hline
\cite{37} & 2012 & - Computational assumptions & - Mutual authentication with key agreement & - Construct an algorithm to solve the CDH problem or the k-CAA problem \\ \hline 
\cite{39} & 2013 & - Pattern recognition approaches & - Continuous authentication & - Feasibility of continuous touch-based authentication \\ \hline 
\cite{41} & 2013 & - Formal proof\newline - Random oracle model & - Transitive authentication & - Solving the CDH problem \\ \hline 
\cite{48} & 2014 & - Game theory & - Anonymous authentication & - Prove that the authentication scheme achieves anonymity, unlinkability, immunity of key-escrow, and mutual authentication \\ \hline 
\cite{68} & 2016 & - Pattern recognition approaches & - Multimodal authentication & - Show that the sensor pattern noise-based technique can be reliably applied on smartphones \\ \hline 
\cite{69} & 2017 & - Pattern recognition approaches & - Active authentication & - Show the performance of each individual classifier and its contribution to the fused global decision \\ \hline \hline
\end{tabular}}
\end{table*}

\begin{figure*}
\centering
	\begin{adjustbox}{width=0.8\textwidth}
		\begin{tikzpicture}[
		Level1/.style   = {rectangle, draw, rounded corners=1pt, thin, align=center, fill=blue!8},
		Level2/.style = {rectangle, draw, rounded corners=1pt, thin, align=center, fill=black!8, text width=8em},
		Level3/.style = {rectangle, draw, rounded corners=1pt, thin, align=left, fill=green!2, 	text width=7.8em},
		level 1/.style={sibling distance=35mm},
		edge from parent path={(\tikzparentnode.south)
			-- +(0,-8pt)
			-| (\tikzchildnode)}]
		
		\node[Level1] {Countermeasures}
		child {node[Level2] (c1) {Cryptographic functions}}
		child {node[Level2] (c2) {Personal identification}}
		child {node[Level2] (c3) {Classification algorithms}}
		child {node[Level2] (c4) {Channel characteristics}};
		
		\begin{scope}[every node/.style={Level3}, node distance=3mm]
		\node [below = of c1, xshift=0.5em] (c11) {\small{Asymmetric encryption function}};
		\node [below = of c11] (c12) {\small{Symmetric encryption function}};
		\node [below = of c12] (c13) {\small{Hash function}};
		
		\node [below = of c2, xshift=0.5em] (c21) {\small{Numbers-based countermeasures}};
		\node [below = of c21] (c22) {\small{Biometrics-based countermeasures}};
		
		\node [below = of c3, xshift=0.5em] (c31) {\small{Logistic regression}};
		\node [below = of c31] (c32) {\small{Naive bayes classifier}};
		\node [below= of c32] (c33) {\small{Decision trees}};
				\node [below= of c32] (c33) {\small{Decision trees}};
		\node [below= of c33] (c34) {\small{Boosted trees
}};
		\node [below= of c34] (c35) {\small{Random forest
}};
		\node [below= of c35] (c36) {\small{Neural networks
}};
		\node [below= of c36] (c37) {\small{Nearest neighbor
}};
		\end{scope}
		
		\foreach \value in {1,2,3}
		\draw[-] (c1.195) |- (c1\value.west);
		
		\foreach \value in {1,...,2}
		\draw[-] (c2.195) |- (c2\value.west);
		
		\foreach \value in {1,...,7}
		\draw[-] (c3.195) |- (c3\value.west);
		
		\end{tikzpicture}
		
	\end{adjustbox}
	\caption{Categorization of countermeasures used by the authentication schemes for smart mobile devices}
\label{fig:Fig1}
\end{figure*}
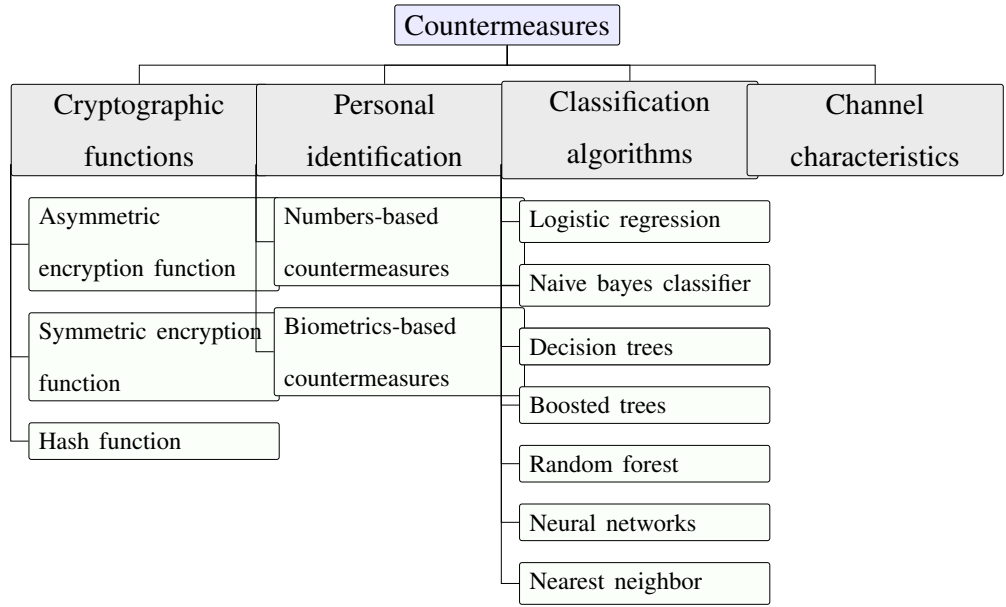

\begin{figure}
\centering
\includegraphics[width=1\linewidth]{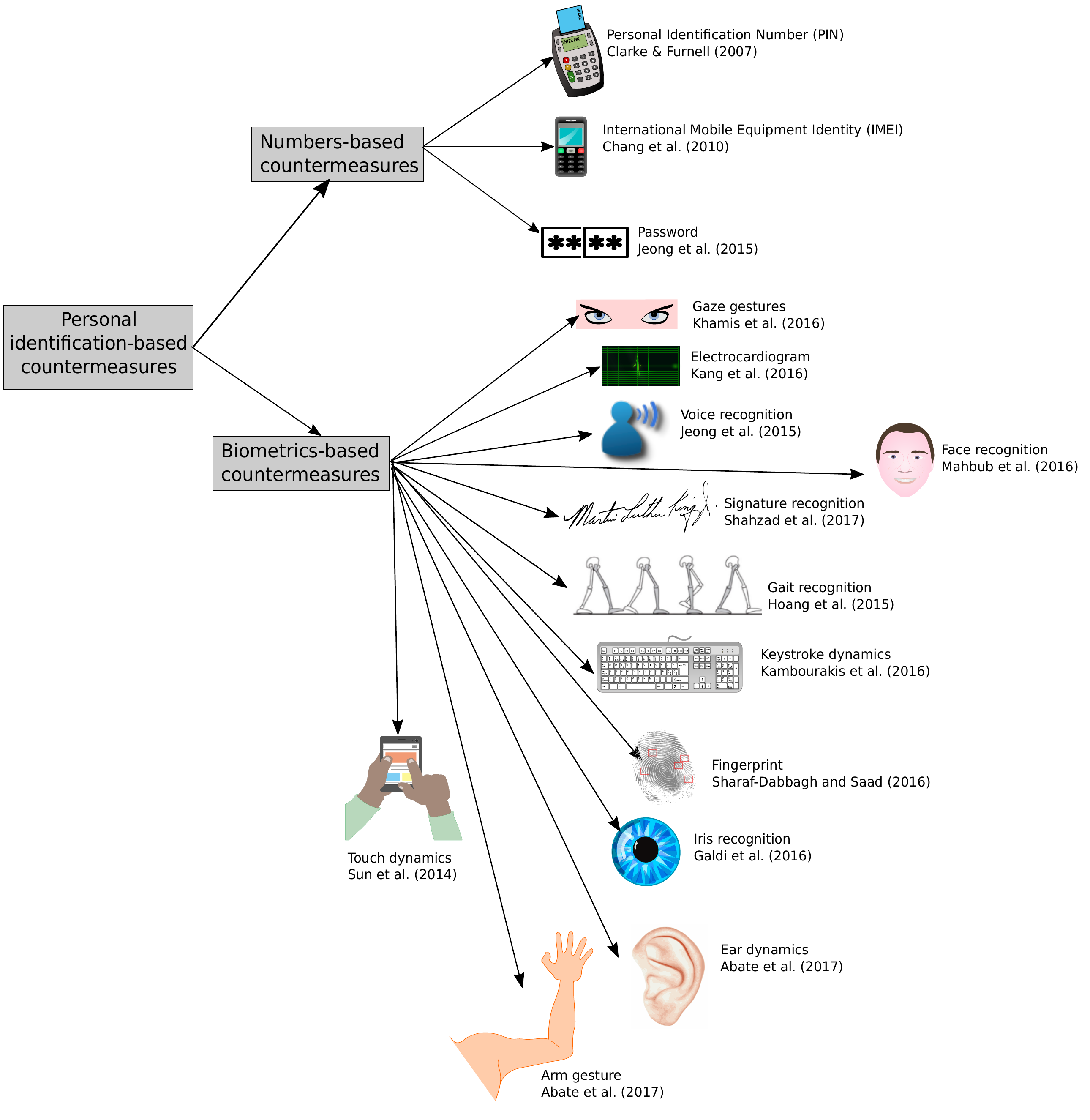}
\caption{Personal identification-based countermeasures used by the authentication schemes for smart mobile devices}
\label{fig:Fig2}
\end{figure}

\begin{figure*}
\centering
\begin{adjustbox}{width=0.9\textwidth}
		\begin{tikzpicture}[
		Level1/.style   = {rectangle, draw, rounded corners=1pt, thin, align=center, fill=blue!7},
		Level2/.style = {rectangle, draw, rounded corners=1pt, thin, align=center, fill=black!8, text width=8em},
		Level3/.style = {rectangle, draw, rounded corners=1pt, thin, align=left, fill=green!2, 	text width=7.8em},
		level 1/.style={sibling distance=37mm},
		edge from parent path={(\tikzparentnode.south)
			-- +(0,-8pt)
			-| (\tikzchildnode)}]
		
		\node[Level1] {Security analysis techniques}
		child {node[Level2] (c1) {Computational assumptions \\~\cite{20,21,23,37}}}
		child {node[Level2] (c2) {Pattern recognition approaches\\~\cite{13,15,33,35,36,39,69,68}}}
		child {node[Level2] (c3) {Formal proof\\~\cite{41}}}
		child {node[Level2] (c4) {Random oracle model\\~\cite{20,23,41}}}
		child {node[Level2] (c5) {Game theory\\~\cite{48}}};
		
		\end{tikzpicture}
		
	\end{adjustbox}
\caption{Categorization of security analysis techniques used by the authentication schemes for smart mobile devices}
\label{fig:Fig3}
\end{figure*}
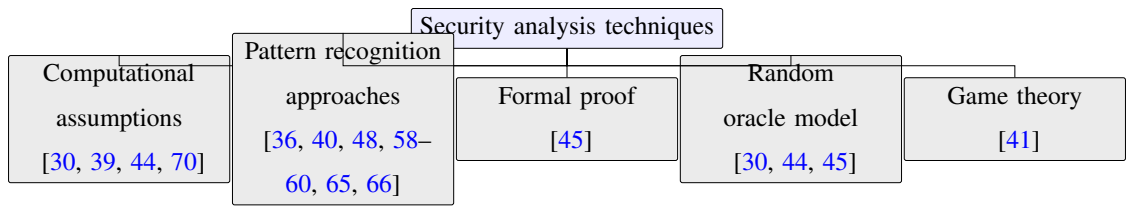

\section{Countermeasures and security analysis techniques}\label{sec:countermeasures-and-security-analysis-techniques}

A secure and efficient authentication scheme is needed to prevent various insider and outsider attacks on many different smart mobile devices. The authentication scheme uses both cryptosystems and non-cryptosystem countermeasures to perform the user authentication whenever a user accesses the devices. In this section, we will discuss the countermeasures and security analysis techniques used by the authentication schemes for smart mobile devices.

\subsection{Countermeasures}

The countermeasures used by the authentication schemes for smart mobile devices can be classified into four categories, including, cryptographic functions, personal identification, classification algorithms, and channel characteristics, as presented in Figure \ref{fig:Fig1}. Table \ref{Table:Tab3}  presents the countermeasures used in authentication schemes for smart mobile devices.

\subsubsection{Cryptographic functions}

The cryptographic functions are used in most authentication schemes for smart mobile devices in order to achieve the security goals, which can be classified into three types of categories, including, asymmetric encryption function, symmetric encryption function and, hash function. As presented in Table \ref{Table:Tab3}, two cryptographic functions are the most used, namely, 1) Bilinear pairings and 2) Elliptic curve cryptosystem (ECC). The authentication schemes \cite{18}, \cite{19}, \cite{21}, \cite{26}, \cite{27}, \cite{104}, \cite{83} use the elliptic curve cryptosystem \cite{119} to reduce the computation loads for mobile devices but they still suffer from some disadvantages such as the need for a key authentication center to maintain the certificates for users' public keys. Using ECC, the scheme \cite{18} provides mutual authentication and supports a session key agreement between the user and the server. The scheme \cite{21} employs ECC and pairing to manipulate authentication parameters and authorization keys for the multiple requests in mobile pay-TV systems. The scheme \cite{26} uses ECC with three-way challenge-response handshake technique in order to provide the agreement of session key and the leaked key revocation capability. Note that hash functions are used specifically to preserve the data integrity. Therefore, in this subsection, we will briefly introduce the bilinear pairings and the elliptic curve cryptosystem.

\paragraph{Bilinear pairings}

Le $G_1$ and $G_2$ be multiplicative groups of the same prime order $p$, respectively. Let $g$ denote a random generator of $G_1$and $e:G_1\times G_1\to G_2$ denote a bilinear map constructed by modified Weil or Tate pairing with following properties:

\begin{itemize}
\item  Bilinear:  $e\left(g^a,g^b\right)={e(g,g)}^{ab},\ \ \forall g\in G_1$ and  $\forall a,\ b\in Z^*_p$. In particular, $Z^*_p=\{x|1\le x\le p-1\}$. 

\item  Non-degenerate: $\exists g\in G_1$ such that $e(g,g)\ne 1$.

\item  Computable: there exists an efficient algorithm to compute $e(g,g)$, $\forall g\in G_1$.
\end{itemize}

\paragraph{Elliptic curve cryptosystem}

As discussed by Guo et al. \cite{49}, the bilinear pairing operations are performed on elliptic curves. An elliptic curve is a cubic equation of the form $y^2+axy+by=x^3+cx^2+dx+e$, where $a$, $b$, $c$, $d$, and $e$ are real numbers. In an elliptic curve cryptosystem (ECC) \cite{119}, the elliptic curve equation is defined as the form of $E_p\left(a,b\right):y^2=x^3+ax+b({\rm mod}\ p)$ over a prime finite field F, where $a,\ b\in F_p$, $p>3$, and $4a^3+27b^2\ne 0\ ({\rm mod}\ p)$. Given an integer $s\in F^*_p$ and a point $P\in E_p(a,b)$, the point multiplication $s\cdot P$ over $E_p(a,b)$ can be defined as $s\cdot P=P+P+\cdots +P$ $(s\ times)$. Generally, the security of ECC relies on the difficulties of the following problems \cite{18}:

\textbf{Definition 1.} Given two points $P$ and $Q$ over $E_p(a,b)$, the elliptic curve discrete logarithm problem (ECDLP) is to find an integer $s\in F^*_p$ such that $Q=s\cdot P$.

\textbf{Definition 2. }Given three points $P$, $s\cdot P$, and $t\cdot P$ over $E_p(a,b)$ for $s,\ t\in F^*_p$, the computational Diffie-Hellman problem (CDLP) is to find the point $(s\cdot P)\cdot P$ over $E_p(a,b)$.\textbf{}

\textbf{Definition 3. }Given two points\textbf{ }$P$ and\textbf{ }$Q=s\cdot P+t\cdot P$ over $E_p(a,b)$ for $s,\ t\in F^*_p$, the elliptic curve factorization problem (ECFP) is to find two points $s\cdot P$ and $t\cdot P$ over $E_p(a,b)$.

\subsubsection{Personal identification}

As shown in Figure \ref{fig:Fig2}, the personal identification can be classified into two types of categories, including:

\paragraph{Numbers-based countermeasures} (e.g, Personal Identification Number (PIN), International Mobile Equipment Identity (IMEI), and Password). Using the inter-keystroke latency, the Clarke and Furnell's scheme \cite{11} classifies the users based upon entering telephone numbers and PINs, where the users are authenticated based upon three interaction scenarios: 1) Entry of 11-digit telephone numbers, 2) Entry of 4-digit PINs, and 3) Entry of text messages. Similar to the scheme \cite{11}, Clarke and Furnell's framework collects the following input data types, 1) Telephone numbers, 2) Telephone area code (5-Digit), 3) Text message, and ) 4-Digit PIN code. According to Wiedenbeck et al. \cite{120}, the numbers-based countermeasures should be easy to remember; they should be random and hard to guess; they should be changed frequently, and should be different for different user's accounts; they should not be written down or stored in plain text. Therefore, the numbers-based countermeasures are vulnerable to various types of attacks such as shoulder surfing.

\paragraph{Biometrics-based countermeasures}are any human physiological (e.g., face, eyes, fingerprints-palm, or ECG) or behavioral (e.g., signature, voice, gait, or keystroke) patterns. As the PIN codes impede convenience and ease of access, the biometrics-based countermeasures are more popular today compared to the numbers-based countermeasures. Some recent smart mobile devices (e.g., iPhone 5S and up and Samsung Galaxy S5 and up) have started to integrate capacitive fingerprint scanners as part of the enclosure. As shown in Figure \ref{fig:Fig2}, we found 12 types of biometrics used as a countermeasure of authentication. Khamis et al. \cite{62} used the \textit{Gaze gestures} for shoulder-surfing resistant user authentication on smart mobile devices. Therefore, Arteaga-Falconi et al. \cite{61} and Kang et al. \cite{74} used the \textit{electrocardiogram} for biometrics authentication based on cross-correlation of the templates extracted. By recognizing the user's voice through the mic, Jeong et al. \cite{57} used the \textit{voice recognition} for user authentication in mobile cloud service architecture. From images captured by the front-facing cameras of smart mobile devices, Mahbub et al. \cite{70} used the \textit{face recognition} for continuous authentication. Based on the behavior of performing certain actions on the touch screens, Shahzad et al. \cite{90} proposed the idea of using \textit{Gestures and Signatures} to authenticate users on touch screen devices. Using gait captured from inertial sensors, Hoang et al. \cite{51} proposed the \textit{Gait recognition} with fuzzy commitment scheme for authentication systems. Based on the way and rhythm, in which the users interacts with a keyboard or keypad when typing characters, Kambourakis et al. \cite{64} introduced the \textit{Keystroke dynamics} for user authentication in smart mobile devices. In addition, Galdi et al. \cite{68} introduced an authentication scheme using \textit{iris recognition} and demonstrated its applicability on smart mobile devices. Finally, based on the idea that the instinctive gesture of responding to a phone call can be used to capture two different biometrics, Abate et al. \cite{79} used the \textit{ear} and \textit{arm} gesture for user authentication in smart mobile devices.

\subsection{Security analysis techniques}

To prove the feasibility of authentication schemes for smart mobile devices in practice, researchers in the security field use the security analysis techniques \cite{99},\cite{Mo5}, which can be categorized into five types, namely, computational assumptions, pattern recognition approaches, formal proof, random oracle model, and game theory, as shown in Figure\ref{fig:Fig3}. Therefore, authentication schemes for smart mobile devices that use security analysis techniques are summarized in Table\ref{Table:Tab4}. Note that the pattern recognition approaches are used especially by biometric-based authentication schemes. More precisely, Clarke and Furnell \cite{13} used the pattern recognition approaches to evaluating the feasibility of utilizing keystroke information in classifying users. Kim and Hong \cite{15} evaluated the feasibility of utilizing together teeth image and voice in terms of the training time per model and authentication time per image. Through the Sensor Pattern Noise (SPN), Galdi et al. showed that the sensor pattern noise-based technique can be reliably applied on smartphones. Therefore, Wu and Tseng \cite{20} used the random oracle model and computational assumptions to show that the proposed scheme is secure against ID attack and an adversary should not know the previous session keys. Finally, Liu et al. \cite{48} used the game theory to prove that the authentication scheme achieves anonymity, unlinkability, immunity of key-escrow, and mutual authentication.

\begin{figure*}
\centering
\begin{adjustbox}{width=0.8\textwidth}
		\begin{tikzpicture}[
		Level1/.style  = {rectangle, draw, rounded corners=1pt, thin, align=center, fill=blue!8},
		Level2/.style = {rectangle, draw, rounded corners=1pt, thin, align=center, fill=black!8, text width=9em},
		Level3/.style = {rectangle, draw, rounded corners=1pt, thin, align=left, fill=green!2, text width=9em},
		level 1/.style={sibling distance=40mm},
		edge from parent path={(\tikzparentnode.south)
			-- +(0,-8pt)
			-| (\tikzchildnode)}]
		\node[Level1] {Authentication schemes for Smart Mobile Devices}
		child {node[Level2] (c1) {Biometric-based authentication}}
		child {node[Level2] (c2) {Channel-based authentication}}
		child {node[Level2] (c3) {Factors-based authentication}}
		child {node[Level2] (c4) {ID-based authentication}};

		\begin{scope}[every node/.style={Level3}, node distance=3mm]
		\node [below = of c1, xshift=0.5em] (c11) {\small{Gaze gestures~\cite{62}}};
		\node [below = of c11] (c12) {\small{Electrocardiogram~\cite{61,74}}};
		\node [below = of c12] (c13) {\small{Voice recognition~\cite{15,43,57}}};
		\node [below = of c13] (c14) {\small{Signature recognition~\cite{90}}};
		\node [below = of c14] (c15) {\small{Gait recognition~\cite{51}}};
		\node [below = of c15] (c16) {\small{Behavior profiling~\cite{58}}};
		\node [below = of c16] (c17) {\small{Fingerprint~\cite{14,27,32,44,75}}};
		\node [below = of c17] (c18) {\small{Smart card~\cite{22}}};
		\node [below = of c18] (c19) {\small{Multi-touch interfaces~\cite{35,36}}};
		\node [below = of c19] (c110) {\small{Graphical password~\cite{40}}};
		\node [below = of c110] (c111) {\small{Face recognition~\cite{46,57,70}}};
		\node [below = of c111] (c112) {\small{Iris recognition~\cite{46,68}}};
		\node [below = of c112] (c113) {\small{Rhythm~\cite{50,54}}};
		\node [below = of c113] (c114) {\small{Capacitive touchscreen~\cite{52}}};
		\node [below = of c114] (c115) {\small{Ear Shape~\cite{79}}};
		\node [below = of c115] (c116) {\small{Arm gesture~\cite{79}}};
		\node [below = of c116, xshift=0.5em] (c117) {\small{Keystroke dynamics~\cite{64}}};
		\node [below = of c117] (c118) {\small{Touch dynamics~\cite{33,45}}};

			\node [below = of c2] (c21) {\small{Physical proximity~\cite{12}}};
		\node [below= of c21] (c22) {\small{Electronic voting~\cite{17}}};
		\node [below = of c22] (c23) {\small{Seamless roaming~\cite{30}}};
		\node [below = of c23] (c24) {\small{Transitive authentication~\cite{41}}};
				\node [below = of c24] (c25) {\small{Attribute-based authentication~\cite{49}}};
				\node [below = of c25] (c26) {\small{User-habit-oriented authentication~\cite{54}}};
				\node [below = of c26] (c27) {\small{Handover authentication~\cite{69}}};
				
		\node [below = of c3, xshift=0.5em] (c31) {\small{Two-factor~\cite{50,45}}};
		\node [below = of c31] (c32) {\small{Three-factor}~\cite{F4}};
		\node [below= of c32] (c33) {\small{Multi-factor}~\cite{57}};

		\node [below = of c4, xshift=0.5em] (c41) {\small{Remote user
authentication~\cite{91,105,14,22,37,38}}};
		\node [below = of c41] (c42) {\small{Multi-server remote user authentication~\cite{38}}};
		\node [below = of c42] (c43) {\small{One-to-many authentication~\cite{67}}};
	
		\end{scope}
		
		\foreach \value in {1,...,18}
		\draw[-] (c1.195) |- (c1\value.west);
		
		\foreach \value in {1,...,7}
		\draw[-] (c2.195) |- (c2\value.west);
		
		\foreach \value in {1,2,3}
		\draw[-] (c3.195) |- (c3\value.west);
		
		\foreach \value in {1,2,3}
		\draw[-] (c4.195) |- (c4\value.west);
		
		\end{tikzpicture}
		
	\end{adjustbox}
\caption{Categorization of authentication schemes for smart mobile devices}
\label{fig:Fig4}
\end{figure*}
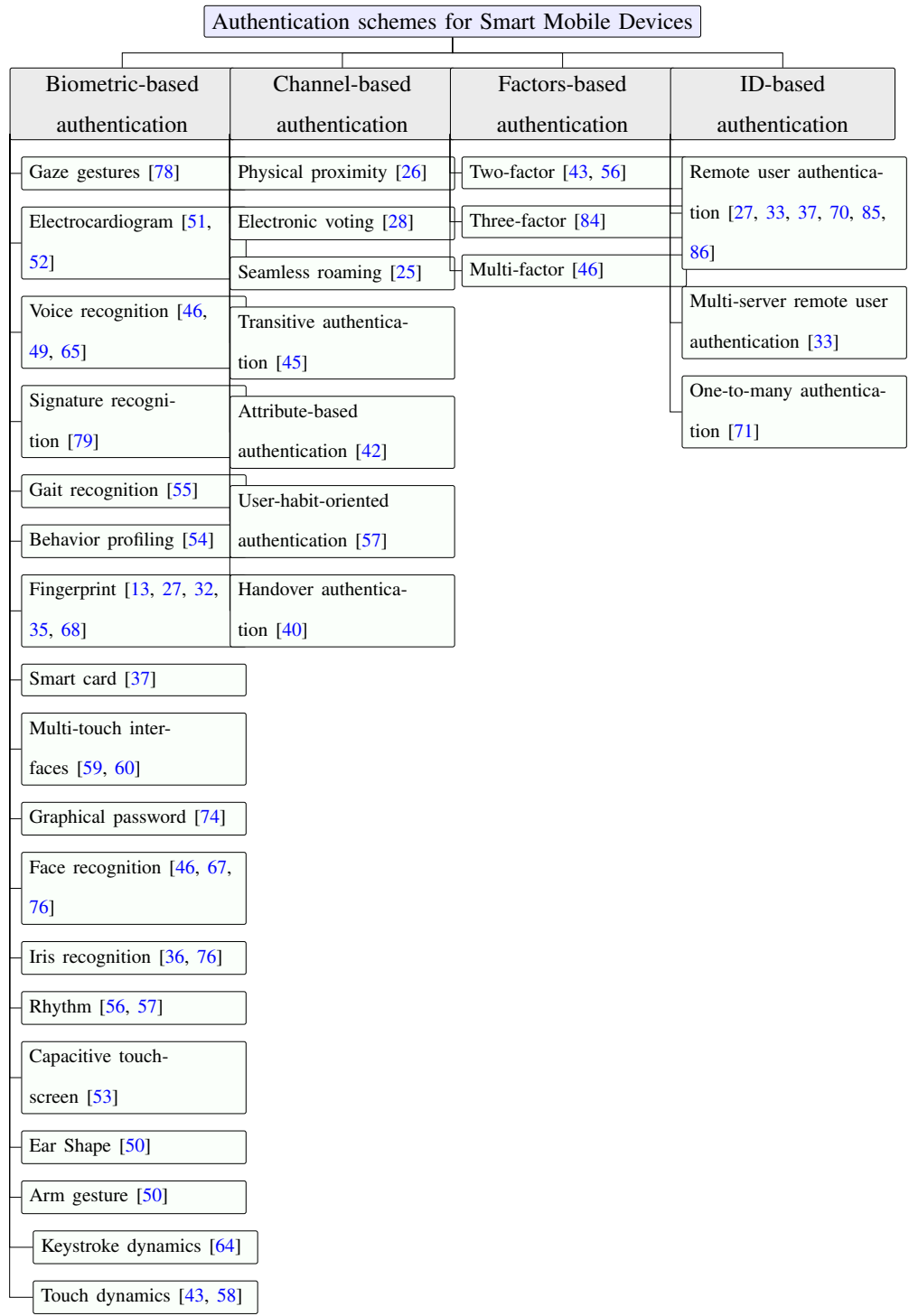
\begin{table}[!t]
	\centering
	\caption{Notations used in comparison of computational costs}
	\label{Table:Tab5}
{\scriptsize
\begin{tabular}{p{0.4in}|p{2.2in}} \hline \hline 
\textbf{Notation} & \textbf{Definition} \\ \hline \hline   
TAR\newline FAR\newline FRR\newline ROC\newline TPR\newline FPR\newline FNR\newline EER\newline GAR\newline $T_e$\newline $T_{mul}$\newline $T_H$\newline $T_{add}$\newline ${TE}_{add}$\newline ${TE}_{mul}$\newline ${TE}_{inv}$\newline $C_1$\newline $C_2$\newline $T_{HE}$ & True acceptance rate\newline False acceptance rate\newline False rejection rate\newline Receiver operating characteristic\newline True-positive rate\newline False-positive rate\newline False-negative rate\newline Equal error rate\newline Genuine acceptance rate\newline  Ttime of executing a bilinear pairing operation\newline  Time of executing a multiplication operation of point\newline Time of executing a one-way hash function\newline Time of executing an addition operation of points\newline Time of executing an elliptic curve point addition\newline Time of executing an elliptic curve point multiplication\newline Time of executing a modular inversion operation\newline Computational cost of client and server (total)\newline Computational cost of subscription (total)\newline Time of encryption and decryption \\ \hline \hline
\end{tabular}
}
\end{table}
\section{Authentication schemes for Smart Mobile Devices}\label{sec:authentication-schemes-for-smart-mobile-devices}

Generally, the classification of authentication schemes frequently mentioned in the literature is done using the following three types, namely, Something-You-Know (can be shared and forgotten), Something-You-Have (can be shared and duplicated), and Someone-You-Are (not possible to share and repudiate), as discussed by Chen et al. in \cite{50, 10}. In our work, according to the countermeasure characteristic used and the authentication model, we categorize the authentication schemes for smart mobile devices in four categories, namely, 1) Biometric-based authentication schemes, 2) Channel-based authentication schemes, 3) Factor-based authentication schemes, and 4) ID-based authentication schemes, as shown in Figure \ref{fig:Fig4}.

\begin{figure}
\centering
\includegraphics[width=1\linewidth]{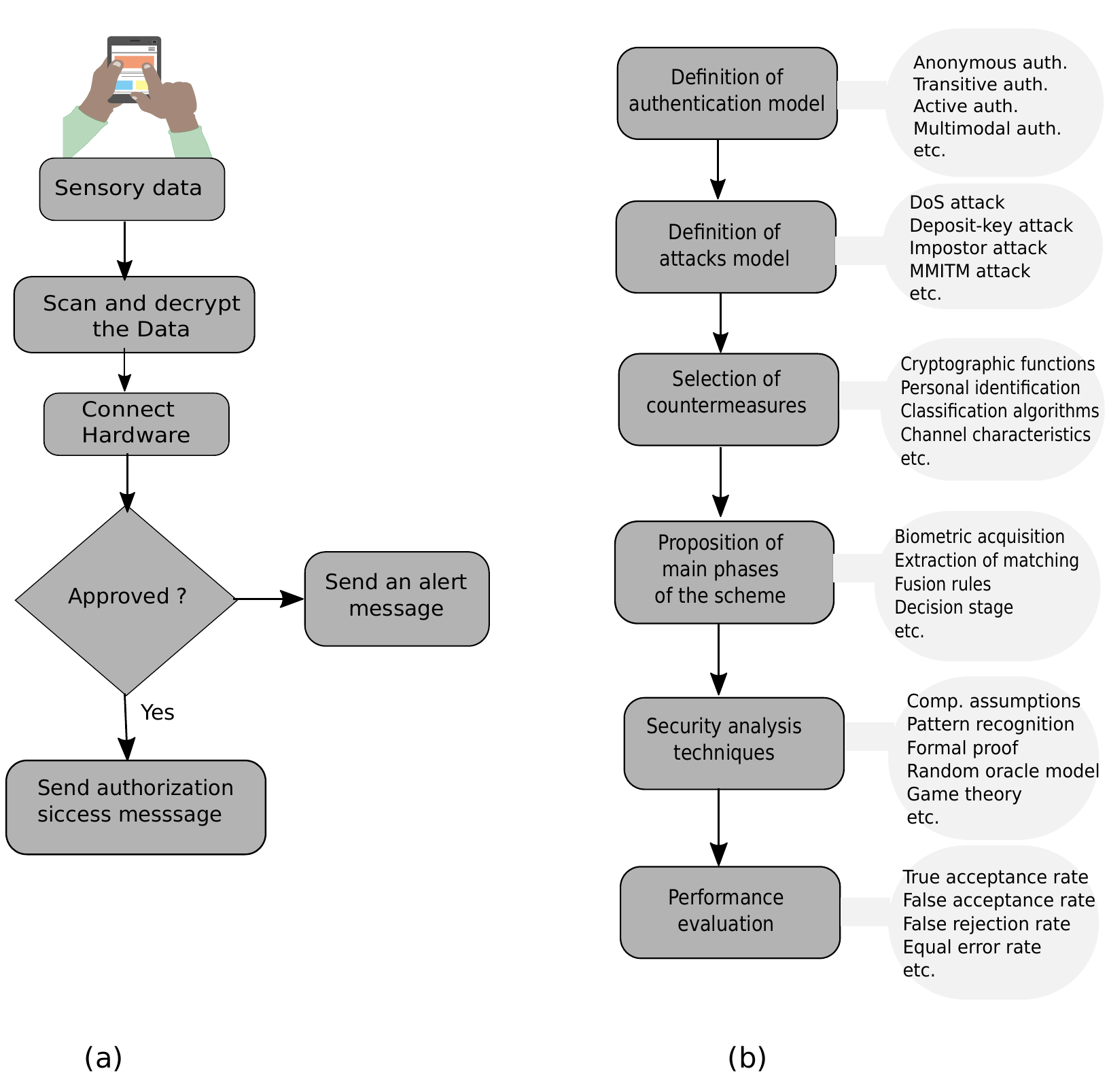}
\caption{Flowcharts depicting the process for (a) authentication using the biometrics-based countermeasures and (b) realization processes of an authentication scheme for smart mobile devices}
\label{fig:Fig5}
\end{figure}

\begin{table*}[!t]
	\centering
	\caption{Biometric-based authentication schemes for smart mobile devices}
	\label{Table:Tab6}
{\scriptsize
\begin{tabular}{p{0.2in}|p{0.6in}|p{0.6in}|p{1.2in}|p{0.8in}|p{1.7in}|p{0.5in}} \hline \hline
\textbf{Time} & \textbf{Scheme} & \textbf{Method} & \textbf{Goal} & \textbf{Mobile device} & \textbf{Performance (+) and limitation (-)} & \textbf{Comp. complexity} \\ \hline \hline
2007 & Clarke and Furnell \cite{11} & - Keystroke analysis & - Introducing the concept of advanced user authentication & - Sony Ericsson T68;\newline - HP IPAQ H5550;\newline - Sony Clie PEG NZ90. & + Keystroke latency\newline - Process of continuous and non-intrusive authentication & Low \\ \hline 
2007 & Clarke and Furnell \cite{13} & - Keystroke analysis & - Enable continuous and transparent identity verification & - Nokia 5110 & + GRNN has the largest spread of performances\newline - The threat model is not defined & High \\ \hline 
2008 & Khan et al. \cite{14} & - Fingerprint & - Introducing the concept of chaotic hash-based fingerprint biometrics remote user authentication scheme & - N/A & + Can prevent from fives attacks, namely, parallel session attack, reflection attack, Forgery attack, impersonation attack, DoS attack, and server spoofing attack\newline - The proposed scheme is not tested on mobile devices & Low \\ \hline 
2010 & Li and Hwang \cite{22} & - Smart card & - Providing the non-repudiation & - N/A & + Can prevent from three attacks, namely, masquerading attacks, replay attacks, and parallel session attacks\newline - Storage costs are not considered & $10T_H$\newline  \\ \hline 
2011 & Xi et al. \cite{27} & - Fingerprint & - Providing the authentication using bio-cryptographic & - Mobile device with Java Platform & + Secure the genuine biometric feature\newline - Server-side attack is not considered & at FAR=0.1\% , GAR=78.69\% \\ \hline 
2012 & Chen et al. \cite{32} & - Fingerprint & - Using only hashing functions & - N/A & + Solve asynchronous problem\newline - Privacy-preserving is not considered & $7T_H$ \\ \hline 
2013 & Frank et al. \cite{39} & - Touchscreen & - Providing a behavioral biometric for continuous authentication & - Google Nexus One & + Sufficient to authenticate a user\newline - Not applicable for long-term authentication & 11 to 12 strokes, EER=2\%--3\% \\ \hline 
2014 & Khan et al. \cite{44} & - Fingerprint & - Improve the Chen et al.'s scheme and Truong et al.'s scheme & - N/A & + Quick wrong password detection\newline - Location privacy is not considered\textbf{} & ${18T}_H$\newline  \\ \hline 
2015 & Hoang et al. \cite{51} & - Gait recognition & - Employing a fuzzy commitment scheme & - Google Nexus One & + Efficient against brute force attacks\newline - Privacy model is not defined & Low \\ \hline 
2016 & Arteaga-Falconi et al. \cite{61} & - Electrocardiogram & - Introducing the concept of electrocardiogram-based authentication & - AliveCor & + Concealing the biometric features during authentication\newline - Privacy model is not considered. & TAR=81.82\% and FAR=1.41\% \\ \hline 
2017 & Abate et al. \cite{79} & - Ear Shape & - Implicitly authenticate the person authentication & - Samsung Galaxy S4 smartphone & + Implicit authentication\newline - Process of continuous and non-intrusive authentication & EER=1\%--1.13\% \\ \hline
2018 & Zhang et al. \cite{203} & - Iris and periocular biometrics & - Develop a deep feature fusion network & - N/A & + Requires much fewer storage spaces \newline - The threat model is limited  & EER= 0.60\% \\ \hline \hline

\end{tabular}
}
\end{table*}
\subsection{Biometric-based authentication schemes}

The surveyed papers of biometric-based authentication schemes for smart mobile devices are shown in Table \ref{Table:Tab6}. As shown in Figure \ref{fig:Fig5}, the realization processes of a biometric-based authentication scheme for smart mobile devices are based on the following processes:

\begin{itemize}
\item  Definition of authentication model (anonymous authentication, transitive authentication, active authentication, multimodal authentication, etc.)

\item  Definition of attacks model (DoS attack, Deposit-key attack, impostor attack, MMITM attack, etc.)

\item  Selection of countermeasures (cryptographic functions, personal identification, classification algorithms, channel characteristics, etc.)

\item  Proposition of main phases of the scheme (biometric acquisition, extraction of matching, fusion rules, decision stage, etc.)

\item  Security analysis techniques (computational assumptions, pattern recognition approaches, formal proof, random oracle model, game theory, etc.)

\item  Performance evaluation (true acceptance rate, false acceptance rate, false rejection rate, equal error rate, etc.)
\end{itemize}

The write a text message using a biometric is called keystroke analysis, which can be classified as either static or continuous. To authenticate users based on the keystroke analysis, Clarke and Furnell \cite{11} introduced the concept of advanced user authentication, which is based on three interaction scenarios, namely, 1) Entry of 11-digit telephone numbers, 2) Entry of 4-digit PINs, and 3) Entry of text messages. The scheme \cite{11} can provide not only transparent authentication of the user and continuous or periodic authentication of the user, but it is also efficient in terms of the false rejection rate and false acceptance rate under three type of mobile devices, namely, Sony Ericsson T68, HP IPAQ H5550, and Sony Clie PEG NZ90. To demonstrate the ability of neural network classifiers, the same authors in \cite{13} proposed an authentication framework based on mobile handset keypads in order to support keystroke analysis. The three pattern recognition approaches used in this framework are, 1) Feed-forward multi-layered perceptron network, 2) Radial basis function network, and 3) Generalised regression neural network. Therefore, Maiorana et al. \cite{31} proved that it is feasible to employ keystroke dynamics on mobile phones with the statistical classifier for keystroke recognition in order to employ it as a password hardening mechanism. In addition, the combination of time features and pressure features is proved by Tasia et al. in \cite{47} that is the best one for authenticating users.

The passwords have been widely used by the remote authentication schemes, which they can be easily guessed, hacked, and cracked. However, to overcome the drawbacks of only-password-based remote authentication, Khan et al. \cite{14} proposed the concept of the chaotic hash-based fingerprint biometrics remote user authentication scheme. Theoretically, the scheme \cite{14} can prevent from fives attacks, namely, parallel session attack, reflection attack, Forgery attack, impersonation attack, DoS attack, and server spoofing attack, but it is not tested on mobile devices and vulnerable to biometric template attacks. To avoid the biometric template attack, Xi et al. \cite{27} proposed an idea based on the transformation of the locally matched fuzzy vault index to the central server for biometric authentication using the public key infrastructure. Compared to \cite{24}, \cite{14}, and \cite{27}, Chen et al. \cite{32} proposed an idea that uses only hashing functions on fingerprint biometric remote authentication scheme to solve the asynchronous problem on mobile devices.

The biometric keys have some advantages, namely, 1) cannot be lost or forgotten, 2) very difficult to copy or share, 3) extremely hard to forge or distribute, and 4) cannot be guessed easily. In 2010, Li and Hwang \cite{22} proposed a biometric-based remote user authentication scheme using smart cards, in order to provide the non-repudiation. Without storing password tables and identity tables in the system, Li and Hwang's scheme \cite{22} can resist masquerading attacks, replay attacks, and parallel session attacks. Therefore, the authors did not specify the application environment of their scheme, but it can be applied to smart mobile devices as the network model is not complicated. Note that Li and Hwang's scheme was cryptanalyzed several times. The question we ask here: is it possible to use a graphical password as an implicit password authentication system to avoid the screen-dump attacks? Almuairfi et al. \cite{40} in 2013, introduced an image-based implicit password authentication system, named IPAS, which is based on creating a visualized image of a user's logged answers.

The touch dynamics for user authentication are initialed on desktop machines and finger identification. In 2012, Meng et al. \cite{33} focused on a user behavioral biometric, namely touch dynamics such as touch duration and touch direction. Specifically, they proposed an authentication scheme that uses touch dynamics on touchscreen mobile phones. To classify users,  Meng et al.'s scheme use known machine learning algorithms (e.g., Naive Bayes, decision tree) under an experiment with 20 users using Android touchscreen phones. Through simulations, the results show that  Meng et al.'s scheme reduces the average error rate down to 2.92\% (FAR of 2.5\% and FRR of 3.34\%). The question we ask here: is it possible to use the multi-touch as an authentication mechanism? Sae-Bae et al. \cite{35} in 2012, introduced an authentication approach based on multi-touch gestures using an application on the iPad with version 3.2 of iOS. Compared with  Meng et al.'s scheme \cite{33},  Sae-Bae et al.'s approach is efficient with 10\% EER on average for single gestures, and 5\% EER on average for double gestures. Similar to  Sae-Bae et al.'s approach \cite{35}, Feng et al. \cite{36} designed a multi-touch gesture-based continuous authentication scheme, named FAST, that incurs FAR=4.66\% and FRR= 0.13\% for the continuous post-login user authentication. In addition, the FAST scheme can provide a good post-login access security, without disturbing the honest mobile users, but the threat model is very limited and privacy-preserving is not considered.

In 2016, Arteaga-Falconi et al. \cite{61} introduced the concept of electrocardiogram-based authentication for mobile devices. Specifically, the authors considered five factors, namely, the number of electrodes, the quality of mobile ECG sensors, the time required to gain access to the phone, FAR, and TAR. Before applying the ECG authentication algorithm, the preprocessing stages for the ECG signal pass by the fiducial point detection. The ECG authentication algorithms are based on two aspects: 1) the use of feature-specific percentage of tolerance and 2) the adoption of a hierarchical validation scheme. The results reveal that the algorithm \cite{61} has 1.41\% false acceptance rate and 81.82\% true acceptance rate with 4s of signal acquisition. Note that the ECG signals from mobile devices can be corrupted by noise as a result of movement and signal acquisition type, as discussed by Kang et al. \cite{74}. However, the advantage of using ECG authentication is concealing the biometric features during authentication, but it is a serious problem if the privacy-preserving is not considered.

\begin{table*}[!t]
	\centering
	\caption{Channel-based authentication schemes for smart mobile devices}
	\label{Table:Tab7}
{\scriptsize
\begin{tabular}{p{0.2in}|p{0.6in}|p{0.6in}|p{1.2in}|p{0.8in}|p{1.7in}|p{0.5in}} \hline \hline
\textbf{Time} & \textbf{Scheme} & \textbf{Method} & \textbf{Goal} & \textbf{Mobile device} & \textbf{Performance (+) and limitation (-)} & \textbf{Comp. complexity} \\ \hline \hline
2007 & Varshavsky et al. \cite{12} & - Physical proximity & - Authenticate co-located devices & - N/A\newline  & + Not vulnerable to eavesdropping\newline - The threat model is limited & High \\ \hline 
2008 & Li et al. \cite{17} & - Electronic voting & - Introducing the concept of  a deniable electronic voting authentication in MANETs & - N/A & + Privacy requirement\newline - Many assumptions needed to understand implementation & Medium \\ \hline 
2011 & He et al. \cite{30} & - Seamless roaming & - Authenticate with privacy-preserving & - N/A & + Privacy requirement\newline - The threat model is limited & Medium \\ \hline 
2013 & Chen et al. \cite{41} & - Tripartite authentication & - Establish a conference key securely & - Samsung Galaxy Nexus & + Transitive authentication\newline - Intrusion detection is not considered & Medium \\ \hline 
2014 & Guo et al. \cite{49} & - Attribute-based authentication & - Authenticate with privacy-preserving & - Nexus S & + Anonymity and untraceability\newline - Interest privacy is not considered & High \\ \hline 
2015 & SETO et al. \newline \cite{54} & - User-habit-oriented authentication & - Integrate the habits with user authentication & - Google Nexus 4 & + More usable for people who have better memory for rhythms than for geometric curves\newline - Privacy is not considered & High \\ \hline 
2016 & Yang et al. \cite{69} & - Handover authentication & - Provides user anonymity and untraceability & - N/A & + Access grant and data integrity\newline - Many assumptions needed to understand implementation & Medium \\ \hline 
2017 & Samangouei et al. \cite{78} & - Attribute-based authentication & - Introducing the concept of facial attributes for active authentication & - Google Nexus 5 & + Implemented with low memory usage\newline - Intrusion detection and encryption are not considered & Medium \\ \hline
2018 & Wu et al. \cite{202} & - Private key security & - Provide both secure key agreement and private
key security & - Samsung Galaxy S5 & + Perfect forward secrecy\newline - Intrusion detection is not considered & Low \\ \hline \hline
\end{tabular}
}
\end{table*}
\subsection{Channel-based authentication schemes}

The surveyed papers of channel-based authentication schemes for smart mobile devices are shown in Table \ref{Table:Tab7}. From dynamic characteristics of radio environment, Varshavsky et al. \cite{12} showed that is possible to securely pair devices using the proximity-based authentication. Specifically, the authors proposed a technique to authenticate co-located devices, named, Amigo. The Amigo scheme uses the knowledge of the shared radio environment of devices as proof of physical proximity, which is specific to a particular location and time. Using the Diffie-Hellman key exchange with verification of device co-location, the Amigo scheme does not require user involvement to verify the validity of the authentication and can detect and avoid the eavesdropping attacks such as the impostor attack and the man-in-the-middle attack.  By exploiting physical layer characteristics unique to a body area network, Shi et al. \cite{42} proposed a lightweight body area network authentication scheme, named BANA. Based on distinct received signal strength variations, the BANA scheme adopts clustering analysis to differentiate the signals from an attacker and a legitimate node. The advantage of BANA scheme is that it can accurately identify multiple attackers with the minimal amount of overhead.

As discussed by the work in \cite{17}, supporting group decisions and especially the electronic voting (e-voting) has become an important topic in the field of mobile applications, where the smart mobile devices can be used to make group decisions electronically. To secure e-voting system, Li et al. \cite{17} proposed that an electronic voting protocol with deniable authentication should satisfy the following requirements: completeness, uniqueness, privacy, eligibility, fairness, verifiability, mobility, and deniable authentication. Based on three types of cryptography, namely, 1) asymmetric encryption function, 2) symmetric encryption function, and 3) hash function, the scheme \cite{17} can meet these requirements of a secure e-voting system for application over mobile ad hoc networks. Theoretically, the scheme \cite{17} can prevent four passive and active attacks, namely, man-in-the-middle attack, impersonation attack, replay attack, and eavesdropping attack, but many assumptions needed to understand the implementation in a smart mobile device.

A roaming scenario in wireless networks involves four parties, namely, a roaming user, a visiting foreign server, a home server, and a subscriber. However, He et al. \cite{30} introduced a user authentication scheme with privacy-preserving, named Priauth, for seamless roaming over wireless networks. Based on probabilistic polynomial time algorithms, the Priauth scheme can satisfy the six requirements: (1) server authentication, (2) subscription validation, (3) provision of user revocation mechanism, (4) key establishment, (5) user anonymity, and (6) user untraceability, but the complexity is high when the Priauth scheme authenticates multiple handheld devices in ad-hoc environment. Using a temporary confidential channel, Chen et al. \cite{41} proposed a bipartite and a tripartite authentication protocol to allow multiple handheld devices to establish a conference key securely, which can reduce the bottleneck of running time human's involvements. 

To provide continuous secure services for mobile clients, it is necessary to design an efficient handover protocol that achieves the handover authentication with user anonymity and untraceability, as discussed in the work \cite{69}. Specifically, Yang et al. use the identity-based elliptic curve algorithm for supporting user anonymity and untraceability in mobile cloud computing. To provide the active authentication on mobile devices, Samangouei et al. \cite{78} introduced the concept of facial attributes.
% * <d.ouahid@gmail.com> 2018-03-27T13:24:39.345Z:
% 
% > Yang et al. 
% Need to cite the work
% 
% ^.

\begin{table*}[!t]
	\centering
	\caption{Factors-based authentication schemes for smart mobile devices}
	\label{Table:Tab8}
{\scriptsize
\begin{tabular}{p{0.2in}|p{0.6in}|p{0.6in}|p{1.2in}|p{0.8in}|p{1.7in}|p{0.5in}} \hline \hline
\textbf{Time} & \textbf{Scheme} & \textbf{Method} & \textbf{Goal} & \textbf{Mobile device} & \textbf{Performance (+) and limitation (-)} & \textbf{Comp. complexity} \\ \hline \hline
2008 & Kim and Hong \cite{15} & - Multimodal biometrics & - Authenticate using teeth image and voice & - Hp iPAQ rw6100 & + Better than the performance obtained using teeth or voice individually\newline - The threat model is not defined & High \\ \hline 
2008 & Yu et al. \cite{16} & - Multimodal biometrics & - Introducing the concept of multimodal biometric-based authentication in MANETs & - N/A & + Biosensor costs\newline - Intrusion detection and encryption are not considered & Medium \\ \hline 
2010 & Park et al. \cite{24} & - Multilevel access control & - Control all accesses to the authorized level of database & - N/A & + Flexibility to dynamic access authorization changes\newline - Many assumptions needed to understand implementation & $10T_H$\newline  \\ \hline 
2012 & Chang et al. \cite{34} & - Graphical password \newline - KDA system & - Combine a graphical password with the KDA system & - Android devices & + Suitable for low-power mobile devices\newline - The threat model is limited & With thumbnails=3, FRR(\%)=7.27, FAR(\%)=5.73 \\ \hline 
2013 & Crawford et al. \cite{43} & - Keystroke dynamics\newline - Speaker verification & - Integrate multiple behavioral biometrics with conventional authentication & - Android devices & + Implement fine-grained access control\newline - No suitable for low-power mobile devices\newline  & Medium \\ \hline 
2014 & Sun et al. \cite{45} & - Multi-touch screens & - Authenticate using multi-touch mobile devices & - Google Nexus 7 & + Robust to shoulder-surfing and smudge attack\newline - Anonymity problem & TPR=99.3\%\newline FPR=2.2\% \\ \hline 
2015 & Chen et al. \cite{50} & - Rhythm & - Authenticate using the rhythm for multi-touch mobile devices & - Google Nexus 7 & + More usable for people who have better memory for rhythms than for geometric curves\newline - Privacy is not considered  & FPR up to 0.7\% \newline FNR up to 4.2\% \\ \hline 
2016 & Khamis et al. \cite{62} & - Gaze gestures\newline - Touch  & - Allow passwords with multiple switches & - Android devices & + Secure against side attacks\newline - The threat model is not defined & Medium \\ \hline 
2016 & Sitova et al. \cite{72} & - Hand movement, orientation, and grasp & - Authenticate using the grasp resistance and grasp stability & - Android devices & + Continuous authentication\newline - Cross-device interoperability & EER=15.1\% \\ \hline 
2017 & Fridman et al. \cite{77} & - Four biometric modalities & - Introducing the active authentication via four biometric modalities & - Android devices & + Active authentication\newline - User reparability & ERR=5\%\newline FRR =1,1\% \\ \hline \hline
\end{tabular}
}
\end{table*}

\begin{figure}
\centering
\includegraphics[width=1\linewidth]{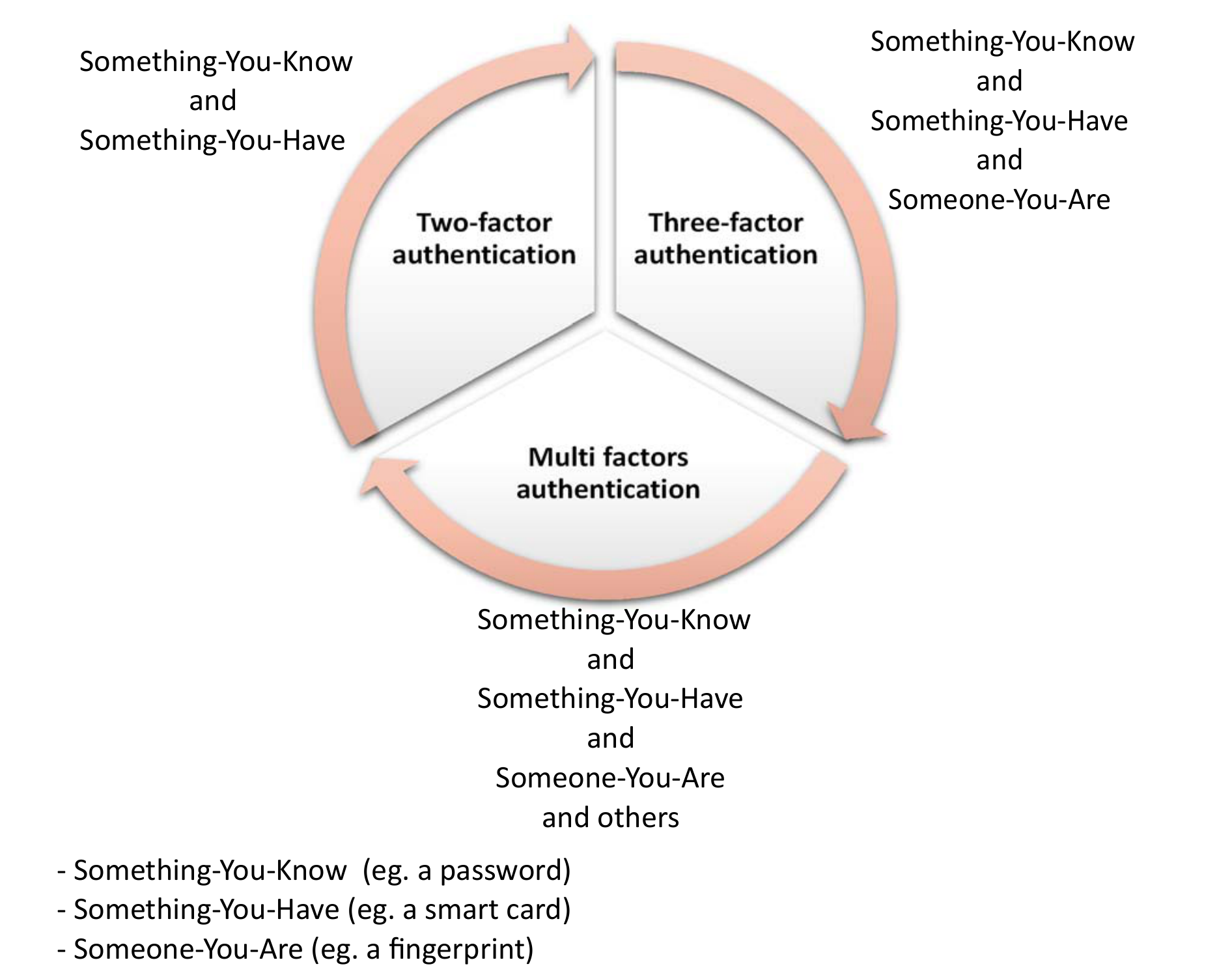}
\caption{Factors-based authentication schemes for smart mobile devices}
\label{fig:Fig7}
\end{figure}

\subsection{Factor-based authentication schemes}

The surveyed papers of factor-based authentication schemes for smart mobile devices are shown in Table \ref{Table:Tab8}. As shown in Figure \ref{fig:Fig7}, factor-based authentication can be classified into three types of categories, including, two-factor authentication, three-factor authentication, and multi-factor authentication.

Kim and Hong \cite{15} proposed a multimodal biometric authentication approach using teeth image and voice. Specifically, this approach is based on two phases, namely, 1) teeth authentication phase and 2) voice authentication phase. The teeth authentication phase uses the AdaBoost algorithm based on Haar-like features for teeth region detection, and the embedded hidden Markov model with the two-dimensional discrete cosine transform. The voice authentication phase uses mel-frequency cepstral coefficients and pitch as voice features. Through performance evolution, the approach was shown that it is better than the performance obtained using teeth or voice individually, but the threat model is not defined. The question we ask here: is it sufficient to use an authentication approach without defining the threat models? Park et al. \cite{24} showed that various attack routes in smart mobile devices may cause serious problems of privacy infringement in data protection. Specifically, using cryptographic methods, the authors designed a combined authentication and multilevel access control, named CAMAC. The CAMAC control uses three types of classification of information level, namely, 1) \textit{Public}, which is not sensitive and can be disclosed in public, 2) \textit{Not public but sharable}, which the data should be encrypted and be decrypted only by authorized users, and 3) \textit{Not public and not sharable}, which the data should be decrypted only by the user himself/herself.

As discussed in the survey \cite{101}, MANET is an autonomous system of mobile nodes (e.g., smart mobile devices), which has several salient characteristics, namely, dynamic topologies, bandwidth constrained and energy constrained operation, and limited physical security. To authenticate the smart mobile devices in MANETs, Yu et al. \cite{16} introduced the concept of multimodal biometric-based authentication, which uses a dynamic programming-based HMM scheduling algorithm to derive the optimal scheme. Therefore, the biosensor scheduling procedure used in the scheme \cite{16} is based on three steps, namely, 1) Scheduling step, to find the optimal biosensor, 2) Observation step, to observe the output of the optimal biosensor and 3) Update step, to judge the result of the authentication. The scheme \cite{16} is efficient in terms of biosensor costs, but the article fails to provide a detailed analysis of intrusion detection and encryption. Related to the scheme \cite{16}, Saevanee et al. \cite{58} proposed a continuous user authentication using multi-modal biometrics with linguistic analysis, keystroke dynamics and behavioral profiling.

Chang et al. \cite{34} proposed the combination of a graphical password with the KDA (Keystroke Dynamic-based Authentication) system for touchscreen handheld mobile devices.  The Chang et al.'s scheme uses the same three phases as in the KDA systems, namely, 1) Enrollment phase, 2) Classifier building phase, and 3) Authentication phase. The enrollment phase is launched when a user's finger presses the touchscreen of the handheld mobile device at thumbnail photo. The classifier building phase is used to verify the user's identity after obtaining the personal features, which the authors employ a computation-efficient statistical classifier proposed by Boechat et al. in \cite{115}. In the authentication phase, the classifier is used to verify the user's identity where the system compares the sequence of graphical password with the registered one in the enrollment phase. Through the experiments, the probability of breaking the Chang et al.'s scheme under a shoulder surfing attack is reduced.

Crawford et al. \cite{43} proposed an extensible transparent authentication framework that integrates multiple behavioral biometrics, namely, keystroke dynamics and speaker verification. The processes of this framework are based on six phases, namely, 1) Update biometric input buffer, 2) Update explicit authentication buffer, 3) Compute individual biometric probability, 4) Compute device confidence, 5) Make task decision, and 6) Update training buffer and refresh classifiers. Therefore, the idea of capacitive touchscreen to scan body parts is proposed by Holz et al. in \cite{51}. Specifically, the authors proposed a biometric authentication system, named Bodyprint, that detects users' biometric features using the same type of capacitive sensing. The Bodyprint system is implemented as an application on an LG Nexus 5 phone, which features a Synaptics ClearPad 3350 touch sensor.

Based on a multimodal recognition of face and iris, De Marsico et al. \cite{46} designed an authentication application, named FIRME, to be embedded in mobile devices. The FIRME is made up of separate modules, with a common starting and final processing, and a central part specialized for each biometrics. The face recognition uses four phases, namely, 1) Acquisition and segmentation, 2) Spoofing detection, 3) Best template selection, and 4) Feature extraction and matching. The iris recognition uses two phases, namely, 1) Acquisition and segmentation and 2) Feature extraction and matching.  The question we ask here is: Is it possible to use the iris liveness detection for mobile devices under the printed-iris attacks? The study published in 2015 by Gragnaniello et al. in \cite{53} proves that with the local binary pattern descriptor, we can detect and avoid the printed-iris attacks using the classification through support vector machine with a linear kernel. Another question we ask here: Is FIRME's scheme effective for the partial face detection? The study published in 2016 by Mahbub et al. in \cite{70} proves that with the fewer facial segment cascade classifiers, we can detect partially cropped and occluded faces captured using a smartphone's front-facing camera for continuous authentication.

The idea of a sequence of rhythmic taps/slides on a device screen to unlock the device is proposed by Chen et al. in \cite{50}.  Specifically, the authors proposed a rhythm-based two-factor authentication, named RhyAuth, for multi-touch mobile devices. The RhyAuth scheme is implemented as an application on Google Nexus 7 tablets powered by Android 4.2. Note that it is possible to use another factor as the third authentication factor such as ID/password. However, the question we ask here is: Is it possible to use four biometric modalities for mobile devices in order to authenticate the users? The study published in 2017 by Fridman et al. in \cite{77} introduced the active authentication via four biometric modalities, namely, 1) text entered via soft keyboard, 2) applications used, 3) websites visited, and 4) physical location of the device as determined from GPS (when outdoors) or WiFi (when indoors).

\begin{table*}[!t]
	\centering
	\caption{ID-based authentication schemes for smart mobile devices}
	\label{Table:Tab9}
{\scriptsize
\begin{tabular}{p{0.2in}|p{0.6in}|p{0.6in}|p{1.2in}|p{0.5in}|p{1.5in}|p{1in}} \hline \hline
\textbf{Time} & \textbf{Scheme} & \textbf{Method} & \textbf{Goal} & \textbf{Mobile device} & \textbf{Performance (+) and limitation (-)} & \textbf{Comp. complexity} \\ \hline \hline
2009 & Yang and Chang \cite{18} & - Elliptic curve cryptosystem & - Providing mutual authentication with key agreement & - N/A & + Resist to outsider, impersonation, and replay attacks\newline - Perfect forward secrecy is not considered compared to the Yoon and Yoo's scheme \cite{19} & ${{3TE}_{mul}+2TE}_{add}$\newline  \\ \hline 
2009 & Yoon and Yoo \cite{19} & - Elliptic curve cryptosystem & - Providing the perfect forward secrecy & - N/A & + Session key security\newline - Location privacy is not considered & ${{1TE}_{mul}+2TE}_{add}$\newline  \\ \hline 
2009 & Wu and Tseng \cite{20} & - Bilinear pairings & - Providing the implicit key confirmation and partial forward secrecy & - N/A & + Secure against a passive attack\newline - The proposed scheme is not tested on mobile devices & $C_1={2T}_e+5T_{mul}+{8T}_H+T_{add}$\newline  \\ \hline 
2009 & Sun and Leu \cite{21} & - Elliptic curve cryptography & - Providing one-to-many facility & - Mobile Pay-TV system\newline  & + Resisting man-in-the-middle attack and replay attack\newline - Interest privacy is not considered & ${C_2=7T}_e+8T_{mul}$ \\ \hline 
2010 & Wu and Tseng \cite{23} & - Bilinear pairings & - Providing the implicit key confirmation and partial forward secrecy & - N/A & + Secure against ID attack\newline - The average message delay and the verification delay are not evaluated & $C_1={2T}_e+6T_{mul}+{6T}_H+2T_{add}$ \\ \hline 
2011 & Islam and Biswas \cite{26} & - Elliptic curve cryptosystem & - Improve the Yang and Chang's scheme  \cite{18} & - N/A & + Prevents user's anonymity problem\newline - Vulnerable to the ephemeral-secret-leakage attacks & ${C_1=4T}_{add}+{8T}_{mul}+7T_H$\newline  \\ \hline 
2012 & He \cite{37} & - Bilinear pairings & - Providing the key agreement and mutual authentication & - HiPerSmart & + Provides key agreement\newline - Perfect forward secrecy is not considered compared to the Yoon and Yoo's scheme \cite{19} & ${C_1=2T}_{add}+{5T}_{mul}+4T_H+T_e+{TE}_{inv}$\newline  \\ \hline 
2013 & Liao and Hsiao \cite{38} & - Self-certified public keys & - Eliminate the risk of leaking the master secret key & - HiPerSmart & + User reparability\newline - Anonymity problem\newline  & ${C_1=2T}_{add}+{10T}_{mul}+7T_H+{2T}_e$ \\ \hline 
2014 & Liu et al. \cite{48} & - Certificateless signature & - Avoiding the forgery on adaptively chosen message attack & - Windows CE 5.2 OS & + Privacy of potential WBAN users\newline - The threat model is limited & $C_1={3T}_e+2T_{mul}+{6T}_H+2T_{add}$ \\ \hline 
2015 & Shahandashti et al. \cite{56} & - Homomorphic encryption & - Achieving implicit authentication & - N/A & + Secure against maliciously-controlled devices\newline - Vulnerable to the replay attack & Medium \\ \hline 
2016 & Islam and Khan \cite{104} & - Elliptic curve cryptosystem & - Providing the user anonymity and unlinkability & - N/A & + Resistance to Pohlig--Hellman attack\newline - Location privacy is not considered & ${C_2=8TE}_{mul}$\newline  \\ \hline
2017 & Wu et al. \cite{83} & - Elliptic curve cryptosystem & - Providing the user anonymity and privacy-preserving & - N/A & + Perfect forward secrecy\newline - Vulnerable to the ephemeral-secret-leakage attacks & $C_1={4TE}_{mul}+{11T}_H$ \\ \hline
2018 & Feng et al. \cite{201} & - Lattice-based anonymous authentication & - Implement an anonymous authentication for the postquantum world & - Samsung GT-I9300 & + Satisfies the identity anonymity and unlinkability characteristics\newline - Interest privacy is not considered & Low \\ \hline \hline
\end{tabular}
} 
\end{table*}

\begin{figure}
\centering
\includegraphics[width=1\linewidth]{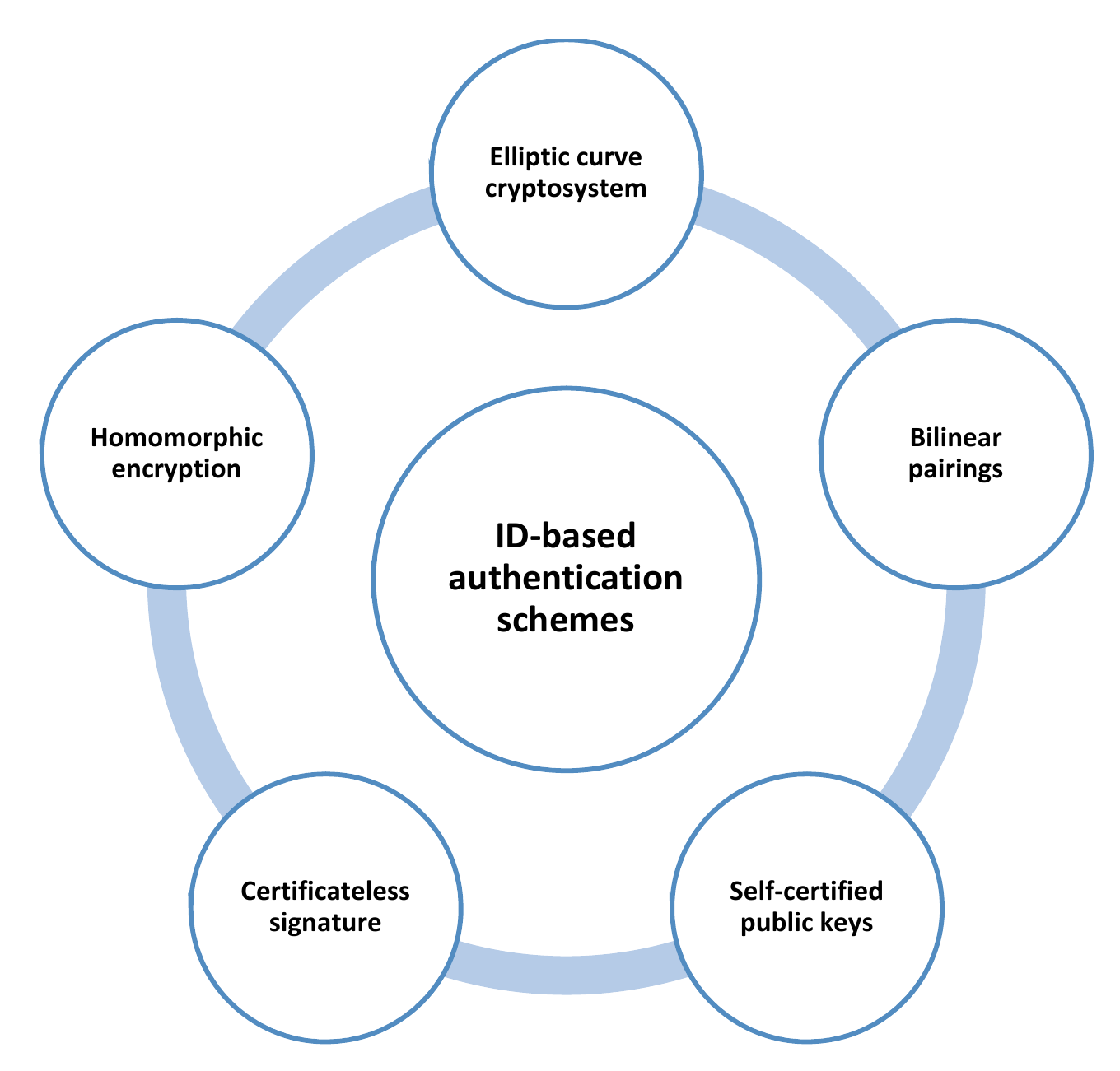}
\caption{Methods used to preserve the authentication models in ID-based authentication schemes for smart mobile devices}
\label{fig:Fig6}
\end{figure}

\begin{table*}
	\centering
	\caption{Summary of Open Research Issues}
	\label{Table:Tab10}
    {\scriptsize
		\begin{tabular}{p{1in}|p{1.3in}|p{1.2in}|p{1in}|p{1.5in}} \hline 
		\textbf{Challenges} & \textbf{Description} & \textbf{Focus/Objective} & \textbf{Contribution} & \textbf{Research opportunities} \\ \hline \hline
		False data injection attacks in mobile cyber-physical system & False data injection attacks jeopardize the system operations in smart mobile devices\textbf{} & How to identify and mitigate false data injection attacks in the mobile cyber-physical system? & Conventional false data detection approaches &  - How to evaluate the overall running status?\newline  - How to design a reputation system with an adaptive reputation updating? \\ \hline 
		Analysis of smart mobile devices under topology attacks & Malicious attacker steals the topology  & How to identify the topology attacks and reduce the amount of stolen information\textbf{} & A stochastic Petri net approach &  - How to proof the efficacy of using a stochastic Petri net approach ?\newline  - How to prove that Petri nets can be useful for modeling mobile cyber-physical system? \\ \hline Integration of smart mobile devices using new generation optical infrastructure technologies (NGN)  & Integration of smart mobile devices with different types of networks such as IoT, vehicular networks, smart grids, ...etc. & How the smart mobile devices are able to mutually authenticate with NGN without any significant increase in overheads ? & An energy-aware encryption for smart mobile devices in Internet of Multimedia Things &  - How to integrate smart mobile devices into NGN ? \newline  - How to design an authentication scheme that reduces the costs in terms of storage cost, computation complexity, communication overhead, and delay overhead? \\ \hline 
		Android malware or malfunctioning smart mobile devices & Malicious or malfunctioning smart mobile devices can be the source of data  & How to safeguard data against such attacks? & An efficient end-to-end security and encrypted data scheme  & - The choice of encryption is a challenge in view of power complexities of smart mobile devices \\ \hline 		Anonymous profile matching & Malicious or malfunctioning smart mobile devices identify a user who has the same profiles & How to provides the conditional anonymity ? &  Prediction-based adaptive pseudonym change strategy & - How to keeps the service overhead of mobile devices very low? \newline - How to achieve the confidentiality of user profiles? \newline - How to resist against the false data injection from the external attacks ?\\ \hline 
		Group authentication and key agreement security under the 5G network architecture & A group of smart mobile devices accessing the 5G network simultaneously cause severe authentication issues & Rethinking the authentication and key agreement protocols in 3G/LTE networks & A group authentication scheme based on Elliptic Curve Diffie-Hellman (ECDH) to realize key forward/backward secrecy & - How to provide privacy and key forward/backward secrecy? \newline - How to resist the existing attacks including redirection, man-in-the-middle, and denial-of-service attacks, etc.\\ \hline 
		Electrocardiogram-based authentication with privacy preservation for smart mobile devices & Privacy preservation in electrocardiogram-based authentication remains a challenging problem since adversaries can find different ways of exploiting vulnerabilities of the electrocardiogram system  & - How to reduce the acquisition time of Electrocardiogram signals for authentication ? \newline - How to achieve privacy preservation and electrocardiogram integrity with differential privacy and fault tolerance?  & - Proposing new privacy-preserving aggregation algorithms \newline - Proposing a new secure handover session key management scheme & - How to resist sensing data link attack? \newline - How to achieve scalability by performing aggregation operations ? \newline - How to improve the TAR and FAR using deep learning? \\ 
        \hline
		 Authentication for smart mobile devices using Software-defined networking (SDN) and network function virtualization (NFV) & The development of network functions using SDN/NFV remains a challenging problem since mobile malware can disrupt the operation of the protocols between the control and data planes, e.g., OpenFlow \cite{Mo1} and ForCES \cite{Mo2} & - How to achieve mutual authentication by adopting both SDN and NFV technologies? & - Proposing new private data aggregation scheme for authentication & - How to secure against malware attack? \newline - How to achieve the computation efficiency? \\ \hline \hline		\end{tabular}}
\end{table*}

\subsection{ID-based authentication schemes}

The surveyed papers of ID-based schemes for smart mobile devices are shown in Table \ref{Table:Tab9}. With the application of cryptography in authentication schemes, smart mobile devices need additional computations, which causes the computation loads and the energy costs of mobile devices to be very high. To solve this problem, researchers proposed several ID-based authentication schemes using elliptic curve cryptosystem (ECC), as discussed in the work \cite{18}. Therefore, as shown in Figure \ref{fig:Fig6}, there are five methods used to provide the authentication models in ID-based authentication schemes for smart mobile devices, namely, bilinear pairings, elliptic curve cryptosystem, self-certified public keys, certificateless signature, and homomorphic encryption.

In order to provide mutual authentication or a session key agreement, Yang and Chang \cite{18} presented an ID-based remote mutual authentication with key agreement scheme. Specifically, the scheme is based on three phases, namely, system initialization phase, user registration phase, and mutual authentication with key agreement phase. Based on the analysis of computational and communication costs, the scheme \cite{18} is efficient compared to Jia et al.'s scheme \cite{105} and can resist outsider, impersonation, and replay attacks. Therefore, Islam and Biswas \cite{26} have analyzed the disadvantage of Yang and Chang's scheme \cite{18} and found that is inability to protect user's anonymity, known session-specific temporary information attack, and clock synchronization problem. 

Yoon and Yoo's scheme \cite{19} showed that Yang and Chang's scheme \cite{18} is vulnerable to an impersonation attack and does not provide perfect forward secrecy. Similar to both Yang and Chang's scheme \cite{18} and  Yoon and Yoo's scheme \cite{19}, Wu and Tseng \cite{20} proposed an ID-based mutual authentication and key exchange scheme for low-power mobile devices. Using the random oracle model and under the gap Diffie--Hellman group, Wu and Tseng's scheme is secure against an ID attack, impersonation attack, and passive attack. The question we ask here is: Will resistance to the impersonation attack give the reliability of an authentication scheme for smart mobile devices? The new study published in 2017 by Spreitzer et al. in \cite{106} proved that the transition between local attacks and vicinity attacks can be increased under the local side-channel attacks, especially in case of passive attacks. Thereby, the local side-channel attacks need to be studied by the authentication schemes for smart mobile devices.

To provide anonymous authentication in mobile pay-TV systems, Sun and Leu \cite{21} proposed an authentication scheme in order to protect the identity privacy. Based on Elliptic curve cryptography (ECC), the Sun and Leu's scheme can manipulate authentication parameters and authorization keys for the multiple requests. Related to the scheme \cite{21}, HE et al. \cite{67} proposed a one-to-many authentication scheme for access control in mobile pay-TV systems. Therefore, using four mechanisms, namely, symmetrical cryptosystem, asymmetrical cryptosystem, digital signature and one-way hash function, Chen's scheme \cite{107} proposed an effective digital right management scheme for mobile devices. Note that Chang et al. \cite{25} have found that Chen's scheme \cite{107} is insecure because an attacker can easily compute the symmetric key, and they proposed an improved schema based on three phases: the registration phase, the package phase, and the enhanced authorization phase.

\section{Open Research Issues}\label{sec:future-directions}

Table \ref{Table:Tab10} summarizes the future directions in authentication issues for smart mobile devices.

\subsection{Android malware or malfunctioning smart mobile devices}
In 2016 \cite{ref1,ref2}, an Android malware succeeded in bypassing the two-factor authentication scheme of many banking mobile apps. The malware can steal the user's login credential, including the SMS verification code. When the legitimate application is launched, the malware is triggered and a fake login screen overlays the original mobile banking one, with no option to close it. After that, the user fills in their personal data in the fake app. The key success of this attack is based on the phishing technique, which displays a graphical user interface (GUI) that has similar visual features as the legitimate app.  The malware can also intercept two-factor authentication code (i.e., verification code sent through SMS), and forward it the attacker. One research direction to prevent this kind of attacks is to detect the apps which have the similar visual appearance and are installed on the same mobile device.

\subsection{Rethinking authentication on smart mobile devices}
Mobile devices are nowadays an essential part of our everyday life and can be integrated with different types of networks such as IoT, vehicular networks, smart grids, ...etc, as they help the user accessing the required resources and information of these networks. This integration requires rethinking the authentication protocols already proposed for mobile devices and considers the new architecture, the new threats, as well as the implementation feasibility in case of resource-constrained devices.

\subsection{Developing more robust containers against sophisticated attacks}
Employees work very often with their mobile devices by using electronic mail, exchange IM messages (instant messaging) or view files directly on the cloud through an online cloud storage application. This means that corporate data is at high risk unless we take the necessary measures to ensure that data are protected and safe. One solution is to secure files with the use of a secure container. Containers isolate user's mobile device and emails are encrypted for protection against third-party access and attachments to emails open in the container, in order to prevent leakage to third-party applications. Future research should focus on developing more robust containers against sophisticated attacks or implementing secure App Wrapping techniques.

\subsection{Securing mobile devices based an unsolvable puzzle}
Recently, University of Michigan was funded for producing a computer that is unhackable \cite{ref3}.  MORPHEUS outlines a new way to design hardware so that information is rapidly and randomly moved and destroyed. The technology works to elude attackers from the critical information they need to construct a successful attack. It could protect both hardware and software. This idea can be the basis for future research for securing mobile devices from attackers.

\subsection{Combined intrusion detection and authentication scheme in smart mobile devices}
Intrusion detection capabilities can be built inside the mobile devices in order to spot real-time malicious behaviors. Such techniques must use combined characteristics and exploit and social network analysis techniques \cite{maglaras2014ocsvm}, in order to cope with zero day attacks and small fluctuations in user behavior.  There are many types of algorithms that may be used to mine audit data on real time, that can be applied to mobile devices. Data mining based IDSs have demonstrated higher accuracy, to novel types of intrusion and robust behaviour \cite{dewa2016data}. 

\subsection{False data injection attacks in mobile cyber-physical system}
False data injection attacks are crucial security threats to the mobile cyber-physical system, where the attacker can jeopardize the system operations in smart mobile devices. Recently, Li et al. in \cite{121} proposed a distributed host-based collaborative detection scheme to detect smart false data injection attacks white low false alarm rate. To identify anomalous measurement data reported, the proposed scheme employs a set of rule specifications. However, how to identify and mitigate false data injection attacks in the mobile cyber-physical system? Hence, false data injection attacks in the mobile cyber-physical system should be exploited in the future.

\subsection{Group authentication and key agreement security under the 5G network architecture}
Based on recent advances in wireless and networking technologies such as Software-defined networking (SDN) and network function virtualization (NFV), 5G will enable a fully mobile and connected society. According to Nguyen et al. \cite{Mo6}, the development of network functions using SDN and NFV will achieve an extremely high data rate. Therefore, a group of smart mobile devices accessing the 5G network simultaneously causes severe authentication issues. In a work published in 2018, Ferrag et al. \cite{99} categorized threat models in cellular networks in four categories, namely, attacks against privacy, attacks against integrity, attacks against availability, and attacks against authentication. How to achieve mutual authentication by adopting both SDN and NFV technologies under these threat models? One possible future direction is to develop a group authentication scheme based on Elliptic Curve Diffie-Hellman (ECDH) to realize key forward/backward secrecy. 

\subsection{Electrocardiogram-based authentication with privacy preservation for smart mobile devices}
Privacy preservation in electrocardiogram-based authentication remains a challenging problem since adversaries can find different ways of exploiting vulnerabilities of the electrocardiogram system. Two questions we ask here: How to reduce the acquisition time of Electrocardiogram signals for authentication? and how to achieve privacy preservation and electrocardiogram integrity with differential privacy and fault tolerance? A possible research direction in this topic could be related to proposing new privacy-preserving aggregation algorithms to resist sensing data link attack.

\section{Conclusions}\label{sec:conclusions}
In this article, we surveyed the state-of-the-art of authentication schemes for smart mobile devices. Through an extensive research and analysis that was conducted, we were able to classify the threat models in smart mobile devices into five categories, including, identity-based attacks, eavesdropping-based attacks, combined eavesdropping and identity-based attacks, manipulation-based attacks, and service-based attacks. In addition, we were able to classify the countermeasures into four types of categories, including, cryptographic functions, personal identification, classification algorithms, and channel characteristics. Regarding the cryptographic functions, the surveyed schemes use three types of cryptographic functions, including, asymmetric encryption function, symmetric encryption function, and hash function.

In order to ensure authentication by the personal identification, the surveyed schemes use two types, including, 1) biometrics-based countermeasures, which are any human physiological (e.g., face, eyes, fingerprints-palm, or ECG) or behavioral (e.g., signature, voice, gait, or keystroke pattern); 2) numbers-based countermeasures (e.g, Personal Identification Number (PIN), International Mobile Equipment Identity (IMEI ), and Password). From security analysis perspective, there are five security analysis techniques used in authentication for smart mobile devices, namely, computational assumptions, pattern recognition approaches, formal proof, random oracle model, and game theory. 

According to the countermeasure characteristic and the authentication model used, we were able to classify the surveyed schemes for smart mobile devices in four categories, namely, biometric-based authentication schemes, channel-based authentication schemes, factor-based authentication schemes, and ID-based authentication schemes. In addition, we presented a side-by-side comparison in a tabular form for each category, in terms of performance, limitations, and computational complexity.

There are still exist several challenging research areas (e.g., false data injection attacks in mobile cyber-physical system, analysis of smart mobile devices under topology attacks, Group authentication and key agreement security under the 5G network architecture, and electrocardiogram-based authentication with privacy preservation\dots etc), which can be further investigated in the near future.

\section*{Conflicts of Interest}
We declare that there are no conflicts of interest regarding the publication of this paper.

\bibliographystyle{spbasic_unsort}
\bibliography{ferragmobile}

\end{document}